\documentclass[aip,jcp,reprint,superscriptaddress]{revtex4-1}
\usepackage{graphicx}
\usepackage{dcolumn}
\usepackage{bm}
\usepackage{amsmath,amssymb,amsthm}
\usepackage{color,epsfig,bigints}

\DeclareMathOperator{\erfc}{erfc}

\begin{document}

\title{Multi-resolution dimer models in heat baths with short-range and long-range interactions} 

\author{Ravinda S. Gunaratne}
\email{gunaratner@maths.ox.ac.uk}
\affiliation{Mathematical Institute, University of Oxford, Radcliffe Observatory Quarter, 
Woodstock Road, Oxford, OX2 6GG, United Kingdom}

\author{Daniel B. Wilson}
\email{wilsond@maths.ox.ac.uk}
\affiliation{Mathematical Institute, University of Oxford, Radcliffe Observatory Quarter, 
Woodstock Road, Oxford, OX2 6GG, United Kingdom}

\author{Mark B. Flegg}
\email{mark.flegg@monash.edu}
\affiliation{School of Mathematical Sciences, Monash University,  9 Rainforest walk, 
Clayton campus, Victoria 3168, Australia}

\author{Radek Erban}
\email{erban@maths.ox.ac.uk}
\affiliation{Mathematical Institute, University of Oxford, Radcliffe Observatory Quarter, 
Woodstock Road, Oxford, OX2 6GG, United Kingdom}

\date{\today}

\begin{abstract} 
\vskip 0.1mm
This work investigates multi-resolution methodologies for simulating dimer models. The solvent particles 
which make up the heat bath interact with the monomers of the dimer either through direct collisions (short-range) 
or through harmonic springs (long-range). Two types of multi-resolution methodologies are considered in detail: (a) 
describing parts of the solvent far away from the dimer by a coarser approach; (b) describing each monomer of the 
dimer by using a model with different level of resolution. These methodologies are then utilised to investigate 
the effect of a shared heat bath versus two uncoupled heat baths, one for each monomer. Furthermore the validity of 
the multi-resolution methods is discussed by comparison to dynamics of macroscopic Langevin equations.
\end{abstract}


\maketitle 

\section{Introduction}

\noindent
Molecular dynamics (MD) approaches, based on the rules 
of classical mechanics, have been used to study the 
behaviour of complex biomolecules in biological 
applications~\cite{Rapaport:2004:AMD,Leimkuhler:2015:MD}. 
They are written in terms of the positions and velocities 
of particles, representing either individual atoms or 
groups of atoms, describing parts of a 
biomolecule~\cite{Marrink:2007:MFF,Yeselevskyy:2010:PWM,Riniker:2011:SEP,Darre:2010:ACG}. 
Inter-particle forces in MD models include 
combinations of short-range and long-range 
interactions~\cite{Israelachvili:2011:ISF,Rowlinson:2002:CSH}.
In all-atom MD models, a common example of short-range forces 
are interactions described by the Lennard-Jones 
potential~\cite{LennardJones:1924:DMF,LennardJones:1931:C}, 
while Coulomb forces provide an example of long-range 
forces~\cite{Israelachvili:2011:ISF}. Considering
coarse-grained or caricature MD models, short-range
interaction models include systems when particles
only interact through direct 
collisions~\cite{Holley:1971:MHP,Durr:1981:MMB,Dunkel:2006:RBM,Erban:2014:MDB},
while long-range interactions also include 
models, where particles interact through 
harmonic-springs~\cite{Ford:1965:SMA,Zwanzig:1973:NGL}.
Once the inter-particle interactions are specified, MD describes
the time evolution of the model as a system of ordinary or stochastic 
differential equations for the positions of particles, which can 
also be subject to algebraic constraints, representing bonds
between atoms or fixed internal structures of a 
biomolecule~\cite{Leimkuhler:2015:MD,Bussi:2007:ASU,Frenkel:2002:UMS}.

Biologically relevant simulations have to be done in aqueous 
solutions. A number of water models have been developed in
the literature to use in all-atom MD simulations, including commonly 
used three-site (SPC/E, TIP3P) models~\cite{Huggins:2012:CLW,Mark:2001:SDT}. 
In coarse-grained MD models, water is often treated with the same level 
of coarse-graining as other molecules in the system. For example, 
four water molecules are combined into a single coarse-grained water 
bead in the Martini model~\cite{Marrink:2007:MFF}, while Wat Four
water model~\cite{Darre:2010:ACG} uses four linked beads placed at 
the corners of a tetrahedron to collectively represent 11 water 
molecules. In this paper, we consider two theoretical heat baths 
which enable more analytical progress than solvent models based
on all-atom or coarse-grained water models. In both cases, the
convergence to the Langevin description of the solute particle 
can be established in a certain 
limit~\cite{Holley:1971:MHP,Durr:1981:MMB,Dunkel:2006:RBM,Erban:2014:MDB,Ford:1965:SMA,Zwanzig:1973:NGL}.
Our solute particle will also be treated with the same level 
of simplicity and described as a simple dimer molecule consisting
of two monomers (beads) connected by a spring.

Multi-resolution (hybrid) methods use detailed and coarse-grained 
simulations in different parts of the simulation domain
during the same dynamic simulation~\cite{Flegg:2012:TRM,Robinson:2015:MRS,Praprotnik:2007:ARS,Ensing:2007:ECA}.
Such methods have been developed in different application areas and at different 
spatial and temporal scales in the literature, including dual-resolution 
approaches AdResS and H-AdResS for all-atom MD 
simulations~\cite{Praprotnik:2005:ARM,Potestio:2014:CSS,Zavadlav:2014:ARS,Zavadlav:2015:ARS,Zavadlav:2017:ARS}, 
methods for coupling Brownian dynamics approaches with lattice-based stochastic
reaction-diffusion models~\cite{Flegg:2014:ATM,Flegg:2015:CMC,Robinson:2014:ATM}
or methods which make use of continuum mean-field equations
for the macroscopic component of the simulation~\cite{Smith:2018:SEH,Franz:2013:MRA,Delgado:2009:CAC}.

In some multi-resolution MD approaches, the region of high resolution moves together 
with the large microscopic structure of interest so that the high resolution model
is always used for the whole considered structure, which can range in size from a 
single biomolecule (a protein or a DNA in solution~\cite{Zavadlav:2014:ARS,Zavadlav:2015:ARS})
to virus-like particles~\cite{Machado:2017:MSV,Tarasova:2017:AMD}.
The structure of interest is placed in the centre of the simulation domain and 
it is solvated using a detailed atomistic MD water in its immediate neighbourhood,
which is coupled with a coarse-grained water description in the rest
of the computational domain. 

Another type of multi-resolution modelling is used 
for modelling of macromolecules where a detailed model of an important part of 
a macromolecule is coupled with a coarser model of the rest of
the macromolecule. For example, atomistic detail of the active part of
an enzyme has been coupled with a coarser model of the rest of the 
protein~\cite{Fogarty:2016:MMC}, different resolutions have been used
in bead-spring modelling of DNA~\cite{Rolls:2017:VRR,Rolls:2018:MPB}
or for modelling of polymer melts~\cite{DiPasquale:2014:MTS,DiPasquale:2017:LGD}.

In this paper, we study both multi-resolution approaches using a simple 
dimer model consisting of two monomers (beads) connected by a spring. 
Similar models, where a macromolecule is described as several beads, 
representing parts of the simulated biomolecule, connected by springs, 
have been obtained in the literature using the method of 
ultra-coarse-graining~\cite{Dama:2013:DUC}. Thus our dimer model can be 
considered as a caricature of an ultra-coarse-grained model of a macromolecule. 
We study its behaviour in two theoretical heat baths.
Our investigation focuses on multi-resolution (multiscale) descriptions 
of the solvent which can be described at the microscopic level of 
individual solvent molecules or at the macroscopic (dimer) level 
with the introduction of extrinsic random thermal forces on the monomers. 
We present models of the same dimer with various  
multi-resolution descriptions for the solvent and highlight the conditions 
and reasons, when and why, different model approximations of the solvent 
may be made in simulations.

Our paper is organized as follows. In Section~\ref{sec2}, we introduce 
the macroscopic dimer model with a macroscopic description for solvent forces. 
This macroscopic model is fully described by Langevin 
equations. The Langevin macroscopic model is commonly used in simulation 
due to ease of implementation and analysis. We discuss in Section~\ref{sec2} 
the properties of this description with the intent to use these properties 
as benchmarks against which to compare microscopic and multi-resolution 
solvent models for the same dimer. Two theoretical microscopic approaches to 
model the solvent are introduced and studied through multi-resolution 
(simultaneous microscopic and macroscopic coupled)
modelling in Sections~\ref{sec3} and \ref{sec4}. One of them is based 
on (very) short-range interactions, as heat bath particles only interact 
with the dimer on contact. The other one is at the opposite extreme, 
as it is based on (very) long-range interactions,
where the heat bath is modelled as a system of many harmonic oscillators.

\section{The dimer model}

\label{sec2}

\noindent In this section we will talk exclusively about 
the construction of the model for the dimer which will be used throughout 
this manuscript. In doing so, we describe the solvent at the macroscopic 
level as an extrinsically added random force. The result will be a set 
of Langevin equations. Throughout the manuscript we will modify the treatment 
of the solvent forces at various scales and hybrid resolutions but the 
underlying dimer model will be the same.

We consider a model of a dimer which is described 
by positions of its two monomers, denoted by
${\mathbf X}_1 = [X_{1;1},X_{1;2},X_{1;3}]$ and
${\mathbf X}_2 = [X_{2;1},X_{2;2},X_{2;3}]$,
respectively. Each monomer has the same mass, $M$.
We denote by ${\mathbf R}$ the vector describing
the separation between the monomers, i.e. 
$
{\mathbf R} = {\mathbf X}_2 - {\mathbf X}_1,
$
and by $R$ its magnitude $R = | {\mathbf R} |.$
The interaction between monomers is given in terms
of the potential $\Phi \equiv \Phi(R) : [0,\infty) 
\to {\mathbb R}$, which generates a force on each 
of the monomers with magnitude $\Phi^\prime(R)$.

When the dimer is placed into a heat bath, there are additional 
forces on the two monomers caused by 
interactions with solvent molecules.
The solvent forces can be modelled in a number of different ways and 
at various scales. In this manuscript, we consider two classes of models
to describe the solvent-dimer interactions. The first, 
presented in Section~\ref{sec3}, models the solvent as a bath 
of point particles which collide with the monomers 
and elastic collisions (short-range interactions) contribute 
to the generation of the forces.  In the second case, described in 
Section~\ref{sec4}, solvent molecules are point particles 
which oscillate around and interact at a distance (through long-range 
interactions) with the monomers. The solvent-dimer interactions are 
the sum of harmonic oscillatory forces acting on each of the monomers. 
Importantly, both descriptions under suitable assumptions lead to 
a macroscopic description of the dimer given by the following set 
of Langevin equations 
\begin{eqnarray}
\mbox{d}{\mathbf X}_1 & = & {\mathbf V}_1 \; \mbox{d}t, 
\label{langeqX1}
\\
\mbox{d}{\mathbf V}_1 & = &
\frac{\Phi^\prime(R)}{M} \, \frac{{\mathbf R}}{R}\, \mbox{d}t
- 
\gamma {\mathbf V}_1\, \mbox{d}t 
+ 
\gamma \sqrt{2 D} \; \mbox{d}{\mathbf W}_1, 
\label{langeqV1}
\\
\mbox{d}{\mathbf X}_2 & = & {\mathbf V}_2 \; \mbox{d}t, 
\label{langeqX2}
\\
\mbox{d}{\mathbf V}_2 & = & 
- \frac{\Phi^\prime(R)}{M} 
\, \frac{{\mathbf R}}{R}\, \mbox{d}t
- 
\gamma {\mathbf V}_2 \, \mbox{d}t 
+ \gamma \sqrt{2 D} \; \mbox{d}{\mathbf W}_2, \qquad
\label{langeqV2}
\end{eqnarray}
where 
${\mathbf V}_1 = [V_{1;1},V_{1;2},V_{1;3}]$ 
and
${\mathbf V}_2 = [V_{2;1},V_{2;2},V_{2;3}]$
are velocities of the first and second monomer,
respectively, ${\mathbf W}_1$ and
${\mathbf W}_2$ are three-dimensional vectors of 
independent Wiener processes, 
$D$ is a diffusion coefficient
and $\gamma$ is a friction coefficient, with 
dimension $[\gamma] = [\mbox{time}]^{-1}$.

System (\ref{langeqX1})--(\ref{langeqV2}) provides 
a macroscopic model of the dimer, which we compare 
with microscopic (or multi-resolution) MD simulations which 
explicitly model the solvent. Its validity for 
different MD models can be tested by comparing
values of different dimer's statistics at equilibrium, including 
its expected length $L_d$, dimer velocity autocorrelation 
function $C_d(\tau)$ and dimer diffusion constant $D_d$,
defined by
\begin{eqnarray}
L_d
&=&
\lim_{t \to \infty}
\left\langle
R \right\rangle,
\nonumber
\\
C_d(\tau) &=& 
\lim_{t \to \infty}
\frac{1}{3}
\,
\left\langle \overline{{\mathbf V}}(t+\tau) \cdot \overline{{\mathbf V}}(t) \right\rangle,
\label{vacfdef}
\\
D_d 
&=&
\lim_{t \to \infty}
\frac{1}
{6 t
}
\left\langle
\left(
\overline{{\mathbf X}}(t)
-
\overline{{\mathbf X}}(0)
\right)^2
\right\rangle,
\nonumber
\end{eqnarray} 
where $\overline{{\mathbf X}} = ({\mathbf X}_1 + {\mathbf X}_2)/2$
is the centre of mass of the dimer and 
$\overline{{\mathbf V}} = ({\mathbf V}_1 + {\mathbf V}_2)/2$ is its velocity.
These quantities can be obtained analytically for our macroscopic 
model~(\ref{langeqX1})--(\ref{langeqV2}) as follows. 
Adding equation (\ref{langeqV1}) and equation (\ref{langeqV2}) and noting that 
the sum of two independent Wiener processes is 
another Wiener processes ${\mathbf W}$ 
with an infinitesimal variance which is the sum of the variances of the original two 
processes, we obtain an Ornstein-Uhlenbeck process for $\overline{{\mathbf V}}$
in the following form
\begin{equation*}
\mbox{d}
\overline{{\mathbf V}}
=
- 
\gamma
\overline{{\mathbf V}}
\, \mbox{d}t 
+ 
\gamma \sqrt{D} \; \mbox{d}{\mathbf W}.
\end{equation*}
Therefore, we have
\begin{equation}
C_d(\tau)
=
\frac{D {\hskip 0.35mm} \gamma}{2}
\,
\exp[ - \gamma \tau ].
\label{Cdform}
\end{equation}
Integrating over $\tau$, we deduce
\begin{equation}
D_d
=
\int_0^\infty
C_d(\tau)
\,
\mbox{d} \tau
=
\frac{D}{2}.
\label{Ddvalue}
\end{equation}
Taking the difference of equation~(\ref{langeqV2}) minus equation~(\ref{langeqV1}), 
implementing the over-damped assumption (where $\gamma$ is large) 
and combining the independent Weiner processes into a single Weiner processes 
$\mathbf{W}$ gives
$$
\mathrm{d}\mathbf{R} 
= 
-
\frac{2 \Phi^\prime(R)}{M \gamma} \frac{{\mathbf R}}{R} \, \mathrm{d}t 
+ 
2 \sqrt{D} \, \mathrm{d}\mathbf{W}.
$$
The stationary distribution corresponding to this process is 
proportional to $\exp[-\Phi(R)/(M D {\hskip 0.35mm} \gamma)]$. Normalizing, we find
the distribution of dimer lengths equal to
$$
\varrho(R) 
= 
\frac{\exp\left[- \frac{\Phi(R)}{M D {\hskip 0.3mm} \gamma} \right]}{4\pi
\bigintsss_{\mbox{\scriptsize \raise -0.3mm \hbox{$\,0$}}}^{\mbox{\scriptsize \raise 0.3mm \hbox{$\infty$}}} 
r^2 \exp\left[- \frac{\Phi(r)}{M D {\hskip 0.3mm} \gamma} \right] \, \mathrm{d}r}.
$$  
In the simulations that follow in this manuscript, 
we shall be assuming the dimer potential acts like a linear spring with a rest 
length of $\ell_0$ and a spring constant of $k$ between the two monomers. That is, 
we shall assume 
\begin{equation}
\Phi(R) = \frac{k (R-\ell_0)^2}{2}.
\label{potentialform}
\end{equation}
Each monomer within the dimer is representing a 
half of a molecule of interest and the value of the spring constant
indicates the flexibility in which the molecule can change its shape. 
In this paper, we consider the parameter regime where the spring constant $k$
is sufficiently large so that the dimer has a well-defined structure.
In the limit of large $k$, we have
$\varepsilon = M D {\hskip 0.35mm} \gamma/ (k \, \ell_0^2) \ll 1$. Then, 
$L_d$ can be calculated as 
\begin{equation}
L_d \approx \ell_0 
\left( 1 + \frac{2 M D {\hskip 0.35mm} \gamma}{k \, \ell_0^2} \right), \label{Ldasympt}
\end{equation}
which is valid up to the first order in $\varepsilon$. In particular, the presence of 
heat baths extends the dimer from its rest length on average. In the following 
two sections, we study two theoretical MD models, where we use equations~(\ref{Cdform}),
(\ref{Ddvalue}) and (\ref{Ldasympt}) 
to compare the macroscopic theory with the results obtained by MD simulations.

\section{Short-range interaction heat bath}

\label{sec3}

\noindent
We describe the two monomers as balls with radius $r_0$ and mass 
$M$ which interact with point solvent particles when they collide with them.
In particular, this is a theoretical model of a (very) short-range 
interaction heat bath. Between collisions, monomers follow Newton's
second law of motion in the form
\begin{eqnarray}
M \frac{\mbox{d}{\mathbf V}_1}{\mbox{d}t}
&=&
\Phi^\prime(R) \, \frac{{\mathbf R}}{R} ,
\label{nocol1}
\\
M \frac{\mbox{d}{\mathbf V}_2}{\mbox{d}t}
&=&
- \Phi^\prime(R) \, \frac{{\mathbf R}}{R},
\label{nocol2}
\end{eqnarray}
where, following our notation introduced in Section~\ref{sec2}, positions and
velocities of the monomers are denoted by ${\mathbf X}_i$ and 
${\mathbf V}_i$, respectively, and 
$
{\mathbf R} = {\mathbf X}_2 - {\mathbf X}_1.
$

Our short-range interaction heat bath is described in terms of positions 
${\mathbf x}_i^j$ and velocities ${\mathbf v}_i^j$, of heat bath particles,
where $i=1,2$ is the monomer number and $j=1,2,3,\dots$, is the number
of the heat bath particle. Notice that this formulation allows us
to consider two important cases: (a) each monomer has its own heat
bath; (b) a single heat bath is shared by both monomers. 
By comparing our simulation results in cases (a) and (b),
we can explicitly investigate whether there are any significant 
hydrodynamic interactions between the monomers. In the case (b), we 
simplify our notation by describing particles of the single heat bath by
\begin{equation}
{\mathbf x}^j = {\mathbf x}_1^j = {\mathbf x}_2^j,
\quad
\mbox{and}
\qquad
{\mathbf v}^j = {\mathbf v}_1^j = {\mathbf v}_2^j.
\label{singleheatbath}
\end{equation}
In both cases (a) and (b), we assume that all heat bath particles 
have the same mass, $m$, and define (dimensionless) parameter $\mu$ by 
$$
\mu = \frac{M}{m}.
$$
We are interested in the parameter regime where $\mu \gg 1$. Our MD model 
is based on elastic collisions of heavy  monomers (balls with mass $M$ and radius 
$r_0$) with point heat bath particles with masses $m$. We assume that the 
collisions are without friction, then conservation of momentum and energy 
yields the following formulae for post-collision velocities~\cite{Durr:1981:MMB}
\begin{eqnarray}
\widetilde{\mathbf V}_{i}
&=& 
\left[ {\mathbf V}_{i} \right]^\parallel
+ 
\frac{\mu - 1}{\mu + 1} \, \left[ {\mathbf V}_{i} \right]^\perp
 + 
\frac{2}{\mu + 1} \, \left[{\mathbf v}_i^j \right]^\perp,
\label{elcol1}
\\
\widetilde{\mathbf v}_i^j 
&=& 
\left[ {\mathbf v}_i^j \right]^\parallel
+
\frac{1 - \mu}{\mu + 1} \, \left[{\mathbf v}_i^j \right]^\perp
 + 
\frac{2 \mu}{\mu + 1} \, \left[ {\mathbf V}_{i} \right]^\perp,
\label{elcol2}
\end{eqnarray}
where ${\mathbf v}_i^j$ is the velocity of the heat bath particle
which collided with the $i$-th monomer, tildes denote 
post-collision velocities, superscripts $\perp$ denote projections of 
velocities on the line through the centre of the monomer and the 
collision point on its surface, and superscripts $\parallel$ denote
tangential components. 

Heat bath models based on elastic collisisions (\ref{elcol1})--(\ref{elcol2})
have been studied by a number of 
authors~\cite{Holley:1971:MHP,Durr:1981:MMB,Dunkel:2006:RBM,Erban:2014:MDB}. 
Consider a single monomer in infinite domain $\mathbb{R}^3$, 
and let the heat bath consist of an infinite number of particles with 
positions distributed according to the spatial Poisson process with density
\begin{equation}
\lambda_{\mu} =
\frac{3}{8 r_0^2} \sqrt{\frac{(\mu + 1) \, \gamma}{2 \pi D}}.
\label{lambda3Dexp}
\end{equation}
This means that the number of points in a subset $\Omega$ of $\mathbb{R}^3$ 
has its probability mass function given by the Poisson distribution
with mean $\lambda_\mu |\Omega|$, where $|\Omega|$ is the volume
of $\Omega$. Let the velocities of the heat bath particles be distributed 
according to the Maxwell-Boltzmann distribution 
\begin{equation}
f_\mu({\mathbf v})
= \frac{1}{\sigma_\mu^3(2\pi)^{3/2}} 
\exp \left[ 
- \frac{v_1^2+v_2^2+v_3^2}{2 \sigma_\mu^2} \right],
\label{fvel3Dexp}
\end{equation}
where ${\mathbf v} = [v_1,v_2,v_3]$ and
\begin{equation}
\sigma_\mu = \sqrt{(\mu + 1) \, D \, \gamma }.
\label{sigmai}
\end{equation}
Then the monomer's behaviour is known to converge to the Langevin 
dynamics~\cite{Durr:1981:MMB,Erban:2014:MDB}.
In particular, if we consider that each monomer has its own heat bath, 
we can show that the position and velocity of the monomers, ${\mathbf X}_i$ 
and ${\mathbf V}_i$, converge (in the sense of distributions) to the solution 
of (\ref{langeqX1})--(\ref{langeqV2}) in the limit $\mu \to \infty$.

In reality all beads representing a macromolecule exist within 
a single heat bath. Thus, we ask whether the correlations introduced by 
a bath of solvent which interacts with both monomers has a non-negligible 
affect on the equilibrium statistics of the dimer. Introducing such coupled 
heat baths for both short-range (in this section) and long-range (in Section~\ref{sec4}) 
interactions we study whether there is a significant difference between 
the one-bath and two-bath models as we vary $\ell_0$, the separation distance,
introduced in equation~(\ref{potentialform}). In order to study this problem, 
we make use of multi-resolution modelling.

\subsection{Multi-resolution model using a co-moving frame} 

\label{shramulti}

\begin{figure*}
\centerline{
\epsfig{file=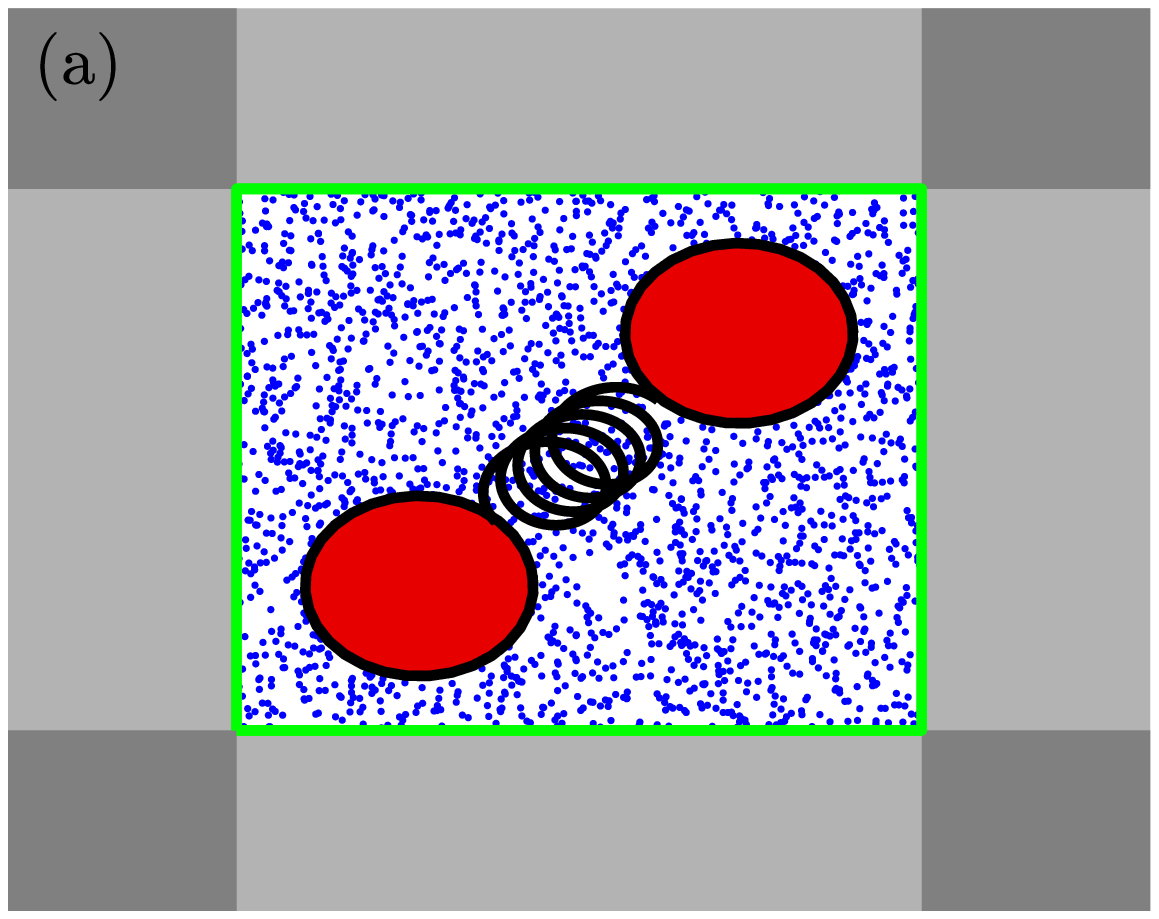,height=5cm}
\hskip -6mm
\epsfig{file=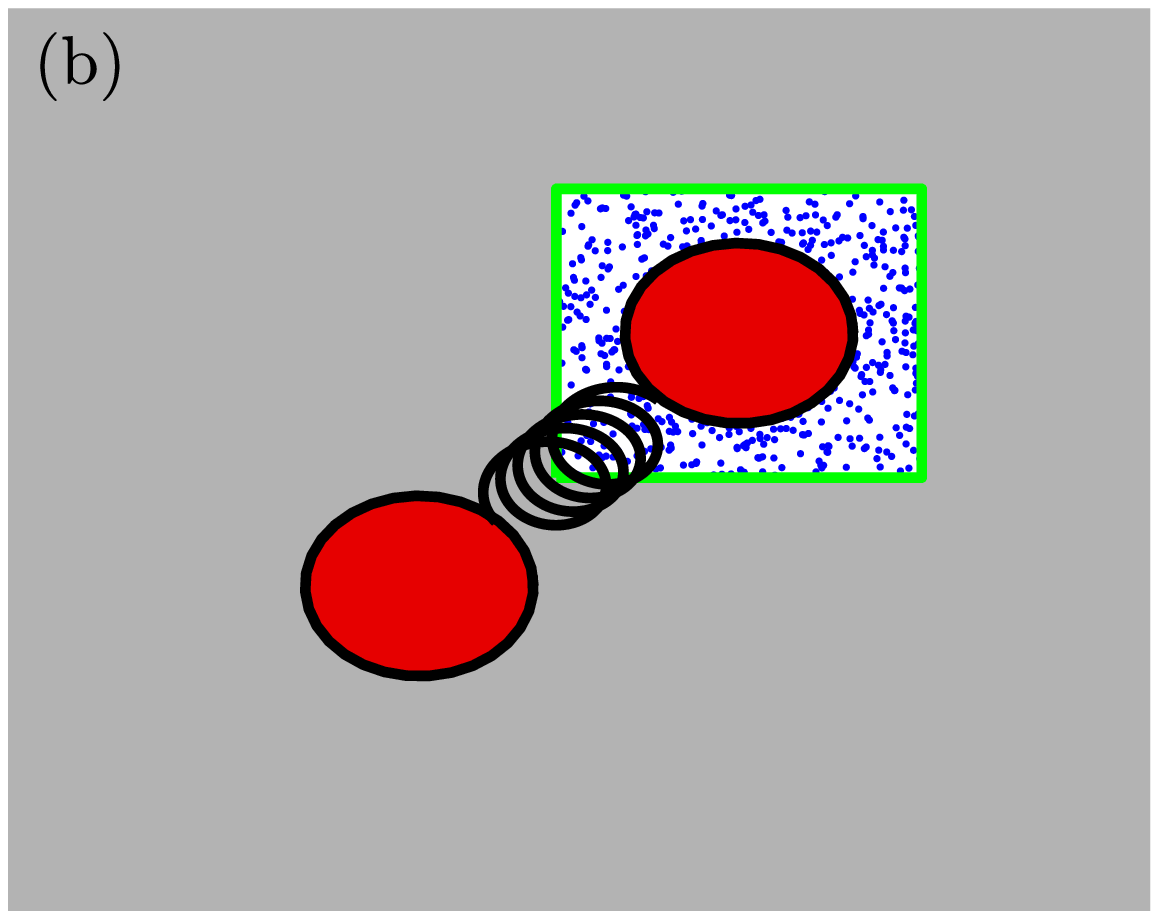,height=5cm}
\hskip -6mm
\epsfig{file=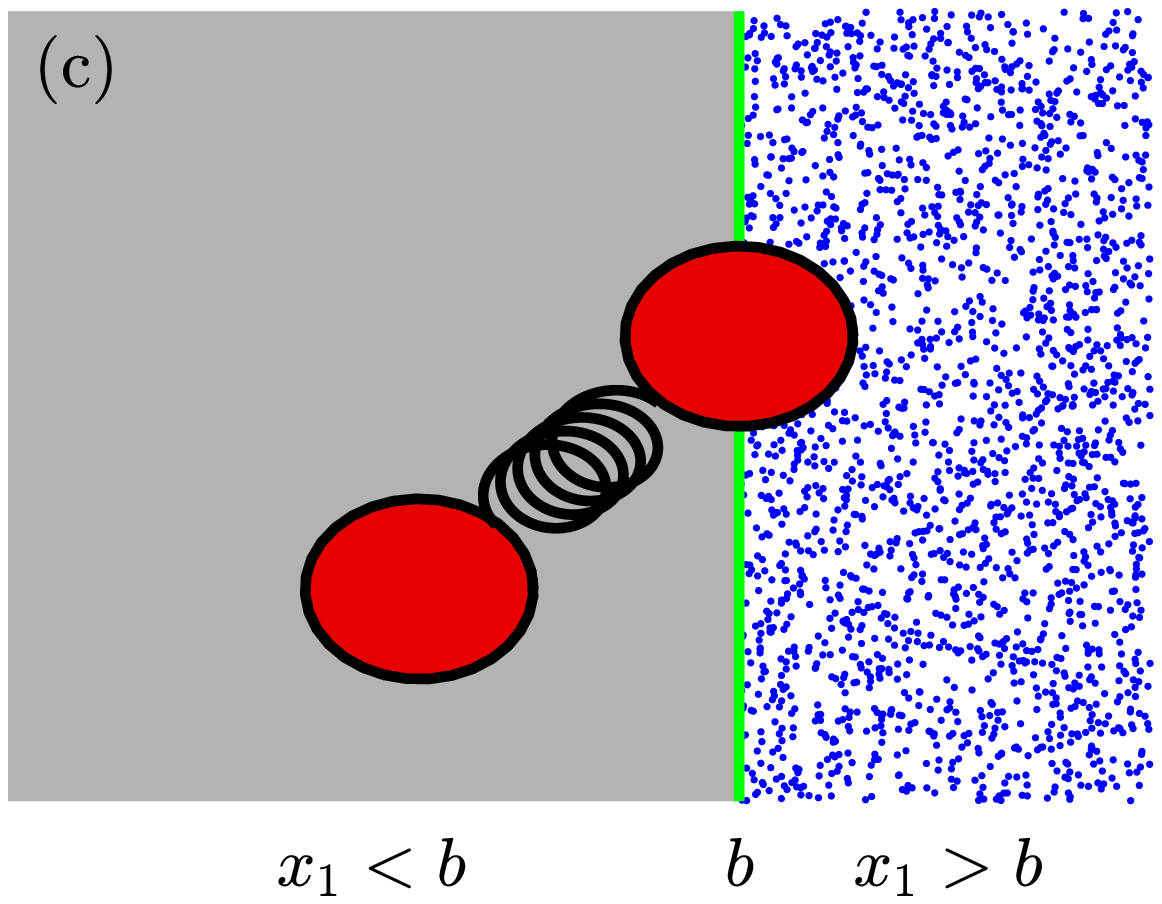,height=5cm}
\hskip 2mm
}
\vskip -0.5cm
\caption{{\it A diagrammatic representation of multi-resolution approaches for a dimer 
in a heat bath with short-range interactions.}
(a) {\it Simulation of the whole dimer in a co-moving frame. The green box depicts the co-moving frame 
that is centred about the dimer. The blue dots correspond to solvent molecules that are explicitly 
modelled. Solvent molecules are not explicitly modelled in the external gray regions.}
(b) {\it Simulation of one monomer in a co-moving frame.}
(c) {\it Simulation with a fixed region of space where an MD model is explicitly used. A dimer molecule
can move to the gray region where it is simulated using the Langevin description.} 
\label{figure1}}
\end{figure*}

\noindent
The solvent in the short-range heat bath interacts with the monomers of the dimer 
through direct contact. In order to simulate the model for long times, 
i.e. where the dimer has undergone a large excursion, the simulated 
domain must be vast as will be the number of solvent particles that must 
be modelled. We present a multi-resolution approach where we only model 
the solvent that is within the close vicinity of the dimer. We consider
a co-moving cubic frame of length $L$ that is centered 
at $\mathbf{X}_{\mbox{\scriptsize f}}(t)$, which we here identify with the centre of mass 
of the dimer at time $t$, i.e. 
\begin{equation}
\mathbf{X}_{\mbox{\scriptsize f}}(t) =
\overline{{\mathbf X}}(t) 
= 
\frac{{\mathbf X}_1(t) + {\mathbf X}_2(t)}{2}.
\label{centreofmasscoframe}
\end{equation}
Within this frame we explicitly model the heat bath with solvent particles,
i.e. they are simulated in the cubic box
\begin{equation}
\mathbf{X}_{\mbox{\scriptsize f}}(t)
+
\left[
-
\frac{L}{2}, 
\frac{L}{2}
\right]
\times
\left[
-
\frac{L}{2}, 
\frac{L}{2}
\right]
\times
\left[
-
\frac{L}{2}, 
\frac{L}{2}
\right].
\label{coframe}
\end{equation}
Externally we model 
the heat bath as a continuum, where the particles are distributed according 
to the spatial Poisson process with density $\lambda_{\mu}$ given in \eqref{lambda3Dexp} 
and the velocities are distributed according to $f_{\mu} ( \mathbf{v} )$ given 
in \eqref{fvel3Dexp}, see Figure~\ref{figure1}(a) for a diagrammatic representation 
of the multi-resolution framework (drawn for clarity in two spatial dimensions,
while all our simulations are three-dimensional). As the dimer moves around 
in $\mathbb{R}^3$ the frame will move with it. In order for the multi-resolution 
model to capture the full model where solvent particles are distributed in the 
entire domain, $\mathbb{R}^3$, we need to introduce new solvent 
particles at the boundary of the frame. 

Consider that time is discretized using small time step $\Delta t$, i.e. 
if the current time is $t$, we want to calculate the state 
of the system at time $t+\Delta t$. In our simulations of the 
multi-resolution model we need the probability of introducing 
a particle at a boundary of frame (\ref{coframe}) in a timestep of length $\Delta t$ 
and subsequently the distribution of the position $\mathbf{x}_{\text{new}}$ 
and velocity $\mathbf{v}_{\text{new}}$ of the new solvent particle. 
For simplicity we transform into the coordinate system 
of the co-moving frame which over an interval of length $\Delta t$ has 
velocity 
\begin{equation}
\textbf{V}_{\mbox{\scriptsize f}} 
= 
\frac{{\mathbf X}_{\mbox{\scriptsize f}}(t+\Delta t) - {\mathbf X}_{\mbox{\scriptsize f}}(t)}{\Delta t}.
\label{framevel}
\end{equation}
The frame 
is always translated to occupy the region $[0,L]^3$. Thus, the velocities for the 
solvent particles in the new reference frame are given by 
$\mathbf{w}^j = \mathbf{v}^j -  \textbf{V}_{\mbox{\scriptsize f}}$. We first calculate the density of particles 
that enter the frame via a particular boundary within a timestep of length $\Delta t$. 
Take, as an illustrative example, the boundary face 
corresponding to $\{x_1=0\}$. Consider particles which
are in half-space $(-\infty,0) \times \mathbb{R}^2$ at time $t$. These
particles have not yet been explicitly included in the simulation. Some of them
will be in half-space $(0,\infty) \times \mathbb{R}^2$ at time $t+\Delta t$.
Their density, $h(x_1)$, only depends on their
first coordinate $x_1 \in (0,\infty)$. We can calculate $h(x_1)$ by
integrating density \eqref{lambda3Dexp}--\eqref{fvel3Dexp}
over solvent particles which are at $x_1^\prime \in (-\infty,0)$ at time $t$ 
and have the appropriate velocity to reach $x_1 \in (0,\infty)$ at time
$t+\Delta t$, namely as~\cite{Erban:2014:MDB}
\begin{eqnarray}
h(x_1) \!&=&\!\!
\int_{-\infty}^0 \int_{\mathbb{R}^2}
\!\!
\lambda_{\mu} 
\,
f_{\mu} \! \left( \! \dfrac{x_1-x_1^\prime}{\Delta t} + V_{\mbox{\scriptsize f};1}, v_2, v_3 \!\right) 
\mathrm{d} v_2 \, \mathrm{d} v_3 \, \mathrm{d} x_1^\prime 
\nonumber
\\
\!&=& \dfrac{\lambda_{\mu}}{2} \,
\text{erfc} \! \left( \dfrac{x_1+V_{\mbox{\scriptsize f};1} \Delta t}{\sigma_{\mu} \Delta t \sqrt{2}}\right), 
\label{eq:errorfunction}
\end{eqnarray}
where $V_{\mbox{\scriptsize f};1}$ is the first component of the frame velocity and 
$\mathrm{erfc}(z) = 2/\sqrt{\pi} \int_z^\infty \exp(-s^2) \, \mbox{{\rm d}}s$ 
is the complementary error function.
Integrating \eqref{eq:errorfunction} over the domain $(0,\infty) \times [0,L] \times [0,L]$
gives us the average number of particles that have entered the frame from the $\{x_1=0\}$ 
boundary in a time interval of length $\Delta t$ as
\begin{eqnarray}
\label{boundary_p}
p_{\mbox{\scriptsize in}} 
&=&
\int_0^\infty \int_0^L \int_0^L h(x_1) \, \mathrm{d} x_3 \, \mathrm{d} x_2 \, \mathrm{d} x_1 
\\
=&& {\hskip -3mm}
\lambda_{\mu} L^2 \Delta t
\Bigg(\!
\dfrac{\sigma_{\mu}}{\sqrt{2\pi}} 
\exp \!\left[ - \dfrac{V_{\mbox{\scriptsize f};1}^2}{2 \sigma_{\mu}^2} \right] 
- 
\dfrac{V_{\mbox{\scriptsize f};1}}{2} 
\erfc \!\left[ \dfrac{V_{\mbox{\scriptsize f};1}}{\sigma_{\mu} \sqrt{2}} \right]
\Bigg). \nonumber
\end{eqnarray}%
In our simulations we choose a timestep small enough that $p_{\mbox{\scriptsize in}} \ll 1$, we can therefore 
use $p_{\mbox{\scriptsize in}}$ as the probability of introducing a new solvent particle. 
Let $\mathbf{z} = [z_1;z_2;z_3]$ be the position of the new solvent 
particle in the coordinate system of the co-moving frame. Then
coordinates $z_2$ and $z_3$ are uniformly distributed in $(0,L)$
and the first coordinate can be sampled from the error function distribution
\begin{equation}\label{eq:posdist}
C_1 \erfc \left[ \dfrac{z_1+V_{\mbox{\scriptsize f};1} \Delta t}{\sigma_{\mu} \Delta t \sqrt{2}} \right], 
\qquad \mbox{for} \;\; z_1 \in (0,\infty),
\end{equation}
where $C_1$ is a normalizing constant. Then the position of the 
new solvent particle in the original coordinates is $\mathbf{x}_{\text{new}} 
= \mathbf{z} + \textbf{X}_{\mbox{\scriptsize f}} (t+ \Delta t) - [L/2,L/2,L/2]$. The velocity, 
$\mathbf{w}$, of the new particle in the co-moving frame must have a first 
coordinate exceeding $z_1 / \Delta t$ in order to reach $z_1$ in a time 
interval of length $\Delta t$. Noting 
that $\mathbf{w} = \mathbf{v}_{\text{new}}-  \textbf{V}_{\mbox{\scriptsize f}}$ 
we write down the distribution of the velocity as the following truncated Gaussian distribution
\begin{equation}\label{eq:veldist}
C_2 \, H \!\left(  v_1 \Delta t - \left( z_1 + V_{\mbox{\scriptsize f};1} \Delta t \right) \right) f_{\mu} (\textbf{v}),
\end{equation}
where $C_2$ is a normalizing constant and $H(\cdot)$ is the Heaviside step function,
satisfying $H(y) = 1$ for $y \in [0,\infty)$ and $H(y)=0$ otherwise.
The position and velocity of solvent particles 
introduced at the other five faces can be done by symmetric modifications of the above 
distributions. 

\begin{figure}
\centerline{\hskip 5mm
\epsfig{file=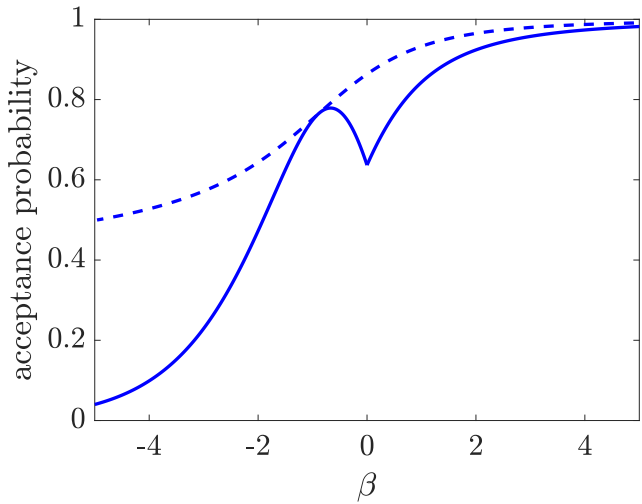,height=7cm}
}
\caption{{\it
Plot of the acceptance probability $(\ref{probacceptance})$ of the 
algorithm presented in Table~{\rm \ref{table1}} for parameters given by $(\ref{choiceofa1})$ 
and $(\ref{choiceofa2})$ (solid line) compared with the acceptance probability
$(\ref{probacceptance})$ calculated for optimal choices of $a_1(\beta)$ and
$a_2(\beta)$ for each parameter value $\beta$.}
\label{figaccprob}}
\end{figure}

Random numbers from distributions \eqref{eq:posdist} and (\ref{eq:veldist}) can be efficiently
sampled using acceptance-rejection algorithms. We use an acceptance-rejection method
for the truncated normal distribution (\ref{eq:veldist}) presented in the
literature~\cite{Robert:1995:STN}, while we sample random numbers from the distribution~\eqref{eq:posdist}
using the acceptance-rejection algorithm presented in Table~\ref{table1}.
\begin{table}

\framebox{%
\hsize=0.961\hsize
\vbox{
\leftskip 6mm
\parindent -4.5mm

\vskip 0.8mm

$\bullet$\hskip 2.53mm
Generate two random numbers $\eta_1$ and $\eta_2$ uniformly distributed in 
interval (0,1).

\smallskip

$\bullet$\hskip 2.53mm
Calculate $a_1(\beta)$ and $a_2(\beta)$ according to (\ref{choiceofa1}) and (\ref{choiceofa2}).

\smallskip

$\bullet$\hskip 2.53mm
Compute an exponentially distributed random number $\eta_3$ by
$\eta_3 = - a_1(\beta) \, \log(\eta_1).$

\smallskip

$\bullet$\hskip 2.53mm
If $\eta_1 \, \eta_2 < a_2(\beta) \, \mathrm{erfc} (\eta_3 + \beta)$,
then choose $\eta_3$ as \hfill\break a sample from the probability distribution
(\ref{shiftederfc}). Otherwise, repeat the algorithm.

\par \vskip 0.8mm}
}\vskip 1mm

\caption{\label{table1} Acceptance-rejection algorithm for sampling random numbers 
according to the probability distribution $p(z;\beta)$ given by~(\ref{shiftederfc}).}
\end{table}%
This is a generalization of the acceptance-rejection algorithm for sampling random numbers according
to the distribution $\sqrt{\pi} \, \mbox{erfc}(z)$ previously used in simulations in the 
stationary frame~\cite{Erban:2014:MDB}. In the case of the distribution \eqref{eq:posdist}, we 
need to sample random numbers according to the probability distribution
\begin{equation} 
p(z;\beta) = C_3(\beta) \, \erfc(z+\beta),
\label{shiftederfc}
\end{equation}
where $\beta \in {\mathbb R}$ is a constant and $C_3(\beta)$ is the normalizing constant
given by
\begin{equation} 
C_3(\beta) = \frac{\sqrt{\pi}}{\exp[-\beta^2]-\sqrt{\pi} \, \beta \, \mbox{erfc}(\beta)}.
\label{shiftederfcnormc}
\end{equation}
The algorithm in Table~\ref{table1} does this by generating an exponentially distributed 
random number $\eta_3$ with mean $a_1(\beta)$, where
\begin{equation}
a_1(\beta)
=
\frac{\sqrt{\pi}}{2}
\times
\left\{
\begin{array}{ll}
\erfc(\beta) \exp(\beta^2),
&
\mbox{ for} \; \beta \ge 0; 
\\
1,
&
\mbox{ for} \; \beta \le 0. 
\\
\end{array}
\right.
\label{choiceofa1}
\end{equation}
To maximise the acceptance probability of this algorithm, we choose its second parameter,
$a_2(\beta)$, as 
\begin{equation}
a_2(\beta)
=
\left\{
\begin{array}{ll}
1/\erfc(\beta),
&
\mbox{ for} \; \beta \ge 0; 
\\
\exp\left(2\beta/\sqrt{\pi}\right),
&
\mbox{ for} \; \beta \le 0. 
\\
\end{array}
\right.
\label{choiceofa2}
\end{equation}
Then its acceptance probability is depending on $\beta$ as
\begin{equation}
\frac{a_2(\beta)}{a_1(\beta) \, C_3(\beta)}.
\label{probacceptance}
\end{equation}
We plot the acceptance probability (\ref{probacceptance})
in Figure~\ref{figaccprob} for our choices 
(\ref{choiceofa1})-(\ref{choiceofa2}) of $a_1(\beta)$ and
$a_2(\beta)$ as the solid line. 

We observe that the acceptance probability (\ref{probacceptance}) for $\beta=0$ is equal to 
$2/\pi \approx 63.7\%$. This value can be improved~\cite{Erban:2014:MDB}
in the case of $\beta = 0$ to 86.3\% provided that we choose
$a_1 = 0.532$ and $a_2 = 0.814$. To obtain a similar improvement
for all values of $\beta$, we could choose both $a_1(\beta)$ and 
$a_2(\beta)$ to maximise the acceptance probability (\ref{probacceptance}),
rather than postulating that $a_1(\beta)$ is given by the piecewise 
defined function (\ref{choiceofa1}) and optimizing $a_2(\beta)$ only. 
The acceptance probability (\ref{probacceptance}) of the resulting
algorithm (which would have $a_1(\beta)$ and $a_2(\beta)$
given by a lookup table, rather than by using formulas 
(\ref{choiceofa1})-(\ref{choiceofa2}))
is plotted in Figure~\ref{figaccprob} as the dashed line 
for comparison.  However, in our illustrative simulations, 
we use the acceptance-rejection algorithm in Table~\ref{table1} with the 
values of $a_1(\beta)$ and $a_2(\beta)$ given by 
(\ref{choiceofa1})-(\ref{choiceofa2}). 

Comparing equations~(\ref{shiftederfc}) 
and~\eqref{eq:posdist}, we observe that we can sample
random numbers from the distribution \eqref{eq:posdist} by sampling
random numbers from the distribution $p(z; V_{\mbox{\scriptsize f};1} \Delta t)$ 
(using the acceptance-rejection algorithm in Table~\ref{table1} 
for $\beta = V_{\mbox{\scriptsize f};1} \Delta t)$) and multiplying them by the
factor $\sigma_{\mu} \Delta t \sqrt{2}$.

One iteration (i.e. an update of the state of the system from time $t$ 
to time $t+\Delta t$) of the multi-resolution simulation algorithm in the co-moving 
frame is given as Algorithm [S1]--[S7] in Table~\ref{table2}. It evolves
the positions and velocities of both monomers together with the positions
and velocities of $N(t)$ solvent particles, where $N(t)$ does depend on time $t$. 
To formulate Algorithm [S1]--[S7], we assume that the timestep $\Delta t$ is chosen 
small enough so that at most one collision happens per iteration. 

We initialize the two monomers with a separation distance $\ell_0$ and generate
a Poisson number (with mean $\lambda_{\mu} \, L^3$) of solvent particles 
in our simulation domain, the cubic frame (\ref{coframe}). The solvent 
particles are initially placed uniformly 
in the frame (\ref{coframe}), where we remove particles overlapping with monomers (before
we begin our simulation) to get the initial number, $N(0)$, of simulated solvent 
particles. Their initial velocities are drawn from the Maxwell-Boltzmann distribution
(\ref{fvel3Dexp}). 

In Step [S1], we update the system over the time interval $(t,t+\Delta t]$ 
using the ``free-flight" positions for each monomer and solvent particle, namely
we use
\begin{eqnarray}
{\widehat{{\mathbf X}}_i}(t+\Delta t) &=& {\mathbf X}_i(t) + {\mathbf V}_i(t) \, \Delta t, 
\label{frfl1} \\
{\widehat{{\mathbf x}}_i^j}(t+\Delta t) &=& {\mathbf x}_i^j(t) + {\mathbf v}_i^j(t) \, \Delta t,
\label{frfl2}
\end{eqnarray}
where $i=1,2$ is the monomer number and $j=1,2,\dots,N(t)$, is the number
of the heat bath particle.
Since $\Delta t$ is chosen so small that only one collision happens during the time interval
$[t,t+\Delta t)$, most of the ``free-flight" positions of solvent particles are accepted 
in Step~[S2] as their updated positions ${\mathbf x}_i^j(t+\Delta t)$ and only the 
solvent particle colliding with a monomer is further updated. 

In Step~[S3], we update the velocities of the monomers by solving (\ref{nocol1})--(\ref{nocol2})
over one time step $[t,t+\Delta t]$. We discretize (\ref{nocol1})--(\ref{nocol2}) using
the forward Euler method as follows
\begin{eqnarray}
{\mathbf V}_1(t+\Delta t) 
&=& 
\widetilde{\mathbf V}_1 + \,
\frac{\Phi^\prime(R)}{M} 
\, \frac{{\mathbf R}}{R}\, \ 
\Delta t, \label{discvelmon1} \\
{\mathbf V}_2(t+\Delta t) 
&=& 
\widetilde{\mathbf V}_2 
- 
\,
\frac{\Phi^\prime(R)}{M} 
\, \frac{{\mathbf R}}{R}\, \ 
\Delta t, \label{discvelmon2} 
\end{eqnarray}
where $\widetilde{\mathbf V}_i$, for $i=1,2,$ is either the post collision velocity 
(if a collision happened in Step~[S2]) or is equal to ${\mathbf V}_i(t)$. In Steps~[S4]--[S5],
we update the position and velocity of the frame. We remove solvent particles which 
are outside of the simulation domain and update $N(t)$ accordingly. 

In Step~[S6],
we use probability $p_{\mbox{\scriptsize in}}$, given by (\ref{boundary_p}), to check whether
any solvent particle entered the simulation domain during the time interval $(t,t+\Delta t]$.
Since $p_{\mbox{\scriptsize in}}$ is the probability of entering the domain through one of its
six sides, we can, for time step  $\Delta t$ small enough that $6 p_{\mbox{\scriptsize in}} \ll 1$,
introduce at most one solvent particle through a randomly chosen side with probability 
$6p_{\mbox{\scriptsize in}}$. 
The initial position and velocity of the introduced solvent particle are sampled according
to distributions \eqref{eq:posdist} and \eqref{eq:veldist} or their symmetric modifications,
taking into account through which side of the cubic frame~(\ref{coframe}) 
the particle entered the frame. 

There is one little
caveat in our derivation of $p_{\mbox{\scriptsize in}}$. To derive equation~\eqref{boundary_p} we 
integrated over the half-space $(-\infty,0) \times \mathbb{R}^2$, meaning that once 
we consider all six faces of the cubic frame~(\ref{coframe}) we have over-counted twice 
at the edges and three times at the corners (as it is highlighted with darker gray
shading in our illustrative diagram in Figure~\ref{figure1}(a)).
This will have negligible effect if we choose $L$ sufficiently large. However, 
it can bias our simulation for values of $L$ comparable with the monomer size $r_0$ 
when $\Delta t$ is not sufficiently small as boundary effects become more pronounced. 
To compensate for this effect, we consider the sampled position,
${\mathbf x}_{\mbox{\scriptsize new}}$ and velocity ${\mathbf v}_{\mbox{\scriptsize new}}$ of the new incoming
particle at time $t+\Delta t$ and calculate its previous position at time $t$ by
$$
{\mathbf y}
=
{\mathbf x}_{\mbox{\scriptsize new}} - {\mathbf v}_{\mbox{\scriptsize new}} \, \Delta t.
$$
If ${\mathbf y}$ is in the regions which were counted twice or three times in
our derivation, we reject the proposed introduction of the 
new solvent particle with the corresponding probability.
Namely, we use the acceptance probability in Step~[S6] given by 
$$
h_{\mbox{\scriptsize acc}}({\mathbf x}_{\mbox{\scriptsize new}},{\mathbf v}_{\mbox{\scriptsize new}})
=
\left\{
\begin{array}{ll}
1,
&
\mbox{ for}\;\; \; {\mathbf y}-\textbf{X}_{\mbox{\scriptsize f}} (t) \in {\mathcal Y}_1;
\\
1/2,
&
\mbox{ for}\;\; \; {\mathbf y}-\textbf{X}_{\mbox{\scriptsize f}} (t) \in {\mathcal Y}_2; 
\\
1/3,
&
\mbox{ for}\;\; \; {\mathbf y}-\textbf{X}_{\mbox{\scriptsize f}} (t)\in {\mathcal Y}_3,
\\
\end{array}
\right.
$$
where ${\mathcal Y}_j \subset {\mathbb R}^3$ is the region of the space
which consists of points which have exactly $j$ of their coordinates outside 
of the interval $[-L/2,L/2]$. For example, in our two-dimensional diagrammatic
representation in Figure~\ref{figure1}(a), the lighter gray shading corresponds 
to region ${\mathcal Y}_1$
while the darker gray shading corresponds to region ${\mathcal Y}_2$.

\begin{table}
\framebox{%
\hsize=0.961\hsize
\vbox{
\leftskip 7.8mm
\parindent -7.8mm

[S1] \hskip 1.2mm
Update the positions of the solvent and the monomers by their ``free-flight" positions
(\ref{frfl1})--(\ref{frfl2}).

\smallskip

[S2] \hskip 1.2mm
If the ``free-flight" position (\ref{frfl2}) of a solvent particle lies within the radius 
of either of the monomers, reverse the trajectories of the solvent and the monomer by time 
$\tau < \Delta t$ such that they are just touching. Calculate post-collision velocities 
by equations~(\ref{elcol1})--(\ref{elcol2}) and update their new 
positions by moving forward by time $\tau$. Otherwise, each ``free-flight" position
is accepted as the particle's position at time $t+\Delta t$.
 
\smallskip

[S3] \hskip 1.2mm
Update the velocities of the monomers by (\ref{discvelmon1})--(\ref{discvelmon2}).

\smallskip

[S4] \hskip 1.2mm
Calculate the new centre of the co-moving frame, $\mathbf{X}_{\mbox{\scriptsize f}}(t+\Delta t)$,
by (\ref{centreofmasscoframe}). Update $N(t)$ by removing solvent particles which 
now lie outside of the frame~(\ref{coframe}) from the simulation.

\smallskip

[S5] \hskip 1.2mm
Calculate the velocity of the frame,
$\textbf{V}_{\mbox{\scriptsize f}}$,  
over the interval $[t,t+\Delta t]$ by equation~(\ref{framevel}). 

\smallskip

[S6] \hskip 1.2mm
Generate two random number $r_1$ and $r_2$ uniformly distributed in interval $(0,1)$. 
If $r < 6p_{\mbox{\scriptsize in}}$, then choose a side of the cube at random and generate proposed position 
${\mathbf x}_{\mbox{\scriptsize new}}$ and velocity ${\mathbf v}_{\mbox{\scriptsize new}}$ of the new solvent 
particle according to distributions \eqref{eq:posdist} and \eqref{eq:veldist}.
If $r_2 < h_{\mbox{\scriptsize acc}}({\mathbf x}_{\mbox{\scriptsize new}},{\mathbf v}_{\mbox{\scriptsize new}})$, then increase
$N(t)=1$ and initialize the new solvent particle at position ${\mathbf x}_{\mbox{\scriptsize new}}$ 
with velocity ${\mathbf v}_{\mbox{\scriptsize new}}$.

\smallskip

[S7] \hskip 1.2mm
Continue with step [S1] using time $t=t+\Delta t$.

\par \vskip 0.8mm}
} 
\caption{\label{table2}
One iteration of the multi-resolution simulation algorithm of the dimer in a co-moving frame.}
\end{table}

\begin{figure}
\centerline{\hskip 5mm
\epsfig{file=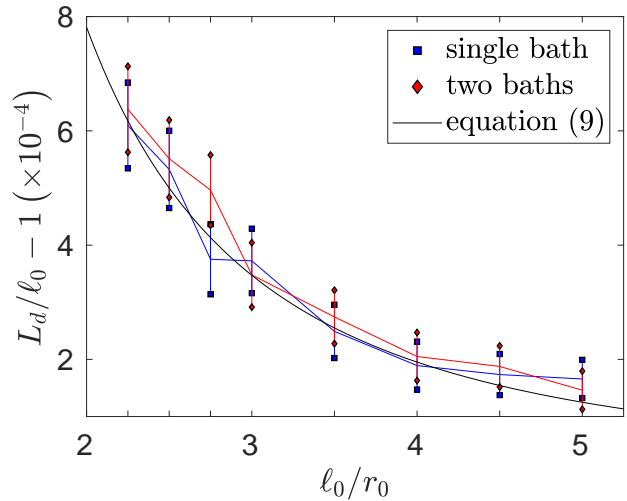,height=7cm}
}
\caption{{\it 
The extension of the average length of a dimer from 
its separation distance $\ell_0$. The equilibrium data for each 
model was collected from a long-time simulation of length $100$
dimensionless time units where $\Delta t = 10^{-6}$ and the
monomers were initially placed with separation $\ell_0$. The values 
of $\alpha=\ell_0/r_0$ presented are $\{2.25,2.5,2.75,3,3.5,4,4.5,5\}$. The 
parameters used are $r_0=0.08$, $\gamma=10$, 
$D=1$, $\mu=10^{3}$, and $k=10^{6}$. In the one-bath case we 
use $L=0.72$ for the frame~$(\ref{coframe})$ enclosing the whole dimer, and for 
the two-bath case, we use $L=0.32$ for each monomer frame.}
\label{figure2}}
\end{figure}

In our illustrative simulations, we use algorithm [S1]--[S7] 
from Table~\ref{table2} together with parameter values 
$r_0=0.08$, $\gamma=10$, $D=1$, $\mu=10^{3}$, $k=10^{6}$ and 
$L=0.72$ for the one-bath case. In Figure~\ref{figure2}, 
we compare simulation results of the average length of 
the dimer at equilibrium, $L_d$, for the one-bath and 
two-bath models. Since the two-bath 
case uses uncoupled heat baths, we can further improve the
efficiency of our algorithm by centering the co-moving frame corresponding to
each heat bath on the corresponding monomer, i.e. we use 
$\mathbf{X}_{\mbox{\scriptsize f}}(t) = \mathbf{X}_i(t)$ for the heat bath
corresponding to the $i$-th monomer in Step~[S4] (instead of the
centre of mass (\ref{centreofmasscoframe})) and choose smaller
value of $L$ in the two-bath case, namely $L=0.32$.
In both one-bath and two-bath models, the solvent particles are distributed 
according to the spatial Poisson process with density $\lambda_\mu$ 
given by~\eqref{lambda3Dexp}. The velocities are distributed according to 
the Maxwell-Boltzmann distribution 
$f_{\mu}( \mathbf{v} )$ given by \eqref{fvel3Dexp}.
We note that in the two-bath case, our model converges to the Langevin dynamics 
\eqref{langeqX1}--\eqref{langeqV2} as $\mu \rightarrow \infty$.
This allows us to attribute any changes 
between the one-bath case and the Langevin model to the correlations induced by 
sharing a heat bath. The asymptotic analytic result obtained for the Langevin model,
equation~(\ref{Ldasympt}), is plotted as the black solid line for comparison.

In Figure~\ref{figure2}, we set the separation distance to be 
$\ell_0 = \alpha \, r_0$ where $\alpha \geq 2$, such that at 
this distance apart the monomers are not overlapping. 
The plot shows the two-sided $99\%$ confidence 
intervals for $(L_d-\ell_0)/\ell_0$ for 
$\alpha \in \{2.25,2.5,2.75,3,3.5,4,4.5,5\}$. Firstly, 
we note that $L_d > \ell_0$ in each of the models as predicted 
in \eqref{Ldasympt}. There seems to 
be reasonable correspondence between the one- and two-bath models, 
with the confidence intervals overlapping. This suggests that the 
correlations we lose by approximating a larger co-moving frame 
around both monomers with two smaller dedicated frames around each
monomer are negligible, allowing us to increase efficiency without 
biasing our overall results. In the next section, we build on
this observation and present a multi-resolution framework which
replaces one of the smaller dedicated frames by a coarser
model of the heat bath, written in terms of the Langevin
dynamics.
  
\subsection{Monomers with different resolution}

\label{secmondifres}

\noindent
As the length of a polymer (i.e. numbers of monomers) increases, a model 
incorporating solvent particles around each of the monomers becomes 
increasingly computationally expensive. However, a fully coarse-grained 
Langevin model of a polymer such as the Rouse model~\cite{Rolls:2017:VRR} 
can lack the required level of detail. Thus, some multi-resolution 
approaches for simulating macromolecules only model an 
important (small) part of a macromolecule using a detailed modelling 
approach~\cite{Fogarty:2016:MMC,Rolls:2017:VRR,Rolls:2018:MPB,DiPasquale:2014:MTS,DiPasquale:2017:LGD}. 
In our case, we can mimic such methodologies by modelling 
the first monomer with explicit solvent with a heat bath of physical 
molecules, while the second monomer is modelled using the Langevin 
equations \eqref{langeqX2} and \eqref{langeqV2}. Such a multi-resolution
approach is schematically shown in Figure~\ref{figure1}(b). To simulate
this model we use a co-moving frame, given by equation (\ref{coframe}),
which is centered around the first monomer, i.e. 
$\mathbf{X}_{\mbox{\scriptsize f}}(t) = \mathbf{X}_1(t)$. 

One iteration of the algorithm is presented as Algorithm [M1]--[M5] 
in Table~\ref{table3}. To begin, we initialize the particle
positions and velocities in the similar way as in
the case of Algorithm [S1]--[S7], with the only difference that the
cubic frame~(\ref{coframe}) is now centered around the first monomer.
Steps [M1] and [M2] are directly equivalent to steps [S1] and [S2].
In Step~[M3], we update the position and velocity of the second 
monomer by
\begin{eqnarray}
{\mathbf V}_2(t+\Delta t) &=& 
{\mathbf V}_2(t) 
- 
\left(
\frac{\Phi^\prime(R)}{M} 
\, \frac{{\mathbf R}}{R}\, \
+
\gamma {\mathbf V}_2(t) \right) 
\Delta t \nonumber \\
& + &
\gamma \, \sqrt{2 D \Delta t} \; \xi, \qquad
\label{langeqV2disc}
\end{eqnarray}
where $\xi$ is sampled from the normal distribution with zero mean and unit
variance. That is, we have replaced the heat bath of the second monomer by solving 
the corresponding Langevin equation (\ref{langeqX1})--(\ref{langeqV2}) using 
the standard Euler-Maruyama integrator. There have been other schemes 
developed in the literature for discretizing the Langevin equation 
such as van Gunsteren and Berendsen~\cite{vanGunsteren:1982:ABD} and 
the Langevin Impulse integrators, which capture the Langevin dynamics 
more accurately especially in the presence of forces, such as the spring 
force between the monomers~\cite{Wang:2003:AFN}. Another 
option would be to consider the BBK integrator~\cite{Brunger:1984:SBC}, 
which we use in Section~\ref{sec4multi}, where we present a multi-resolution 
algorithm for the long-range interaction heat bath model and
discretize the Langevin equation using a combination of the velocity Verlet 
and Euler-Maruyama integrators, see equations~(\ref{velverlet1})-(\ref{langeqA2disc}).
An additional approach is the Verlet scheme~\cite{Gronbech:SEV:2013} that 
approximates the velocity using a central difference discretization 
rather than the forward difference approach used in the Euler-Maruyama 
method, or Runge-Kutta methods~\cite{Burrage:2007:NMS}, which could 
further reduce the error of the multi-resolution simulations.

\begin{table}
\framebox{%
\hsize=0.961\hsize
\vbox{
\leftskip 9mm
\parindent -9mm

[M1] \hskip 1.2mm
Update the positions of the solvent and the monomers by their ``free-flight" positions
(\ref{frfl1})--(\ref{frfl2}).

\smallskip

[M2] \hskip 1.2mm
If the ``free-flight" position (\ref{frfl2}) of a solvent particle lies within the radius 
of the first monomer, reverse the trajectories of the solvent and the monomer by time 
$\tau < \Delta t$ such that they are just touching. Calculate post-collision velocities 
by equations (\ref{elcol1})--(\ref{elcol2}) for $i=1$ and update their new 
positions by moving forward by time $\tau$. Otherwise, each ``free-flight" position
is accepted as the particle's position at time $t+\Delta t$.
 
\smallskip

[M3] \hskip 1.2mm
Update the velocity of the first monomer by~(\ref{discvelmon1}) and 
the velocity of the second monomer by~(\ref{langeqV2disc}).

\smallskip

[M4] \hskip 1.2mm
Calculate the new centre of the co-moving frame as $\mathbf{X}_{\mbox{\scriptsize f}}(t+\Delta t) = \mathbf{X}_1(t)$. 
Update $N(t)$ by removing solvent particles which now lie outside of the frame~(\ref{coframe}) 
from the simulation. Use steps [S5]--[S6] from the algorithm in Table~\ref{table2} to introduce
new solvent particles into the co-moving frame~(\ref{coframe}).

\smallskip

[M5] \hskip 1.2mm
Continue with step [M1] using time $t=t+\Delta t$.

\par \vskip 0.8mm}
} 
\caption{\label{table3}
One iteration of the multi-resolution simulation algorithm of 
the dimer in the heat bath with short-range interactions, 
where the second monomer is simulated by the Langevin dynamics.}
\end{table}

\begin{figure}
\centerline{\hskip 5mm
\epsfig{file=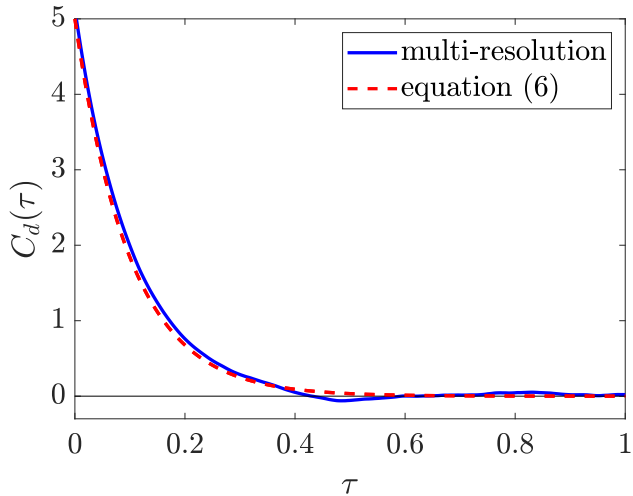,height=7cm}
}
\caption{{\it The velocity autocorrelation 
function for the multi-resolution model (blue solid line) 
for short-range interactions. The function is estimated 
from long time simulation over dimensionless 
time of $500$ time units. It is compared with the result
for the Langevin description of the whole dimer, given
by equation $(\ref{Cdform})$ (red dashed line).
The parameters are
$r_0=0.08$, $\gamma=10$, $D=1$, $\mu=10^{3}$, $k=10^{6}$, 
$\ell_0 = 4r_0$, $\Delta t = 10^{-6}$ and $L=0.32$.}
\label{figure3}}
\end{figure}

In order to compare simulations of the multi-resolution model with 
simulations of the Langevin model  
\eqref{langeqX1}--\eqref{langeqV2} we use the velocity autocorrelation
function of the dimer, $C_d(\tau)$, given by
equation~(\ref{vacfdef}). It has been analytically calculated for the 
Langevin description in equation~(\ref{Cdform}).
In Figure~\ref{figure3}, we present numerical estimates of the 
velocity autocorrelation function of the multi-resolution model from long 
time simulation data, using definition~(\ref{vacfdef}). 

Our results compare well with the theoretical 
result for the Langevin model, though it seems like there is 
a slightly raised value for $C_d(0)$. Using (\ref{Ddvalue}), 
we can estimate the diffusion constant of the dimer $D_d$ by
numerically integrating the velocity auto-correlation function
in interval $[0,1]$. We obtain $D_d \approx 0.529$, while
its theoretical value for the dimer model is given in equation
(\ref{Ddvalue}) as $D/2 = 0.5$. Another approach is to fit 
the exponential function, in the form equation (\ref{Cdform}),
to the computational result presented in Figure~\ref{figure3}.
In this way, the values of both $D$ and $\gamma$ can be estimated 
simultaneously. We found that $D \approx 1.0714$, which is 
higher than our parameter value $D=1$, and $\gamma \approx 9.6064$,
which is lower than $\gamma=10$ used in our simulations. This could suggest 
that the value of $\lambda_\mu$ is too low or that of $\sigma_{\mu}$ 
is too high in our simulations. However, when these quantities 
are measured during the simulations we do not observe any 
deviation. This suggests that, rather than our sampling methods, 
there are small errors introduced by our implementation of the 
moving frame, or more profound boundary effects introduced by the 
small size of the frame. A potential problem in the implementation 
of the co-moving frame, is that solvent particles that leave the 
frame never return. For a stationary frame this is valid as the 
monomer cannot interact with a particle that leaves. 
However, for a co-moving small frame centred about the monomer, 
a solvent particle could leave the frame and return at a later 
time in the simulation. This is not taken into account 
in the presented algorithms.

\section{Long-range interaction heat bath}

\label{sec4}

\noindent
Coarse-grained models of molecular systems can be written in terms of beads 
interacting through coarse-grained force fields. Each bead represents 
a collection of atoms and a coarse-grained potential energy can be constructed
from detailed all-atom MD. Such an approach can usually provide a good description 
of equilibrium properties of molecular systems, but it does not necessarily 
lead to correct dynamics if the time evolution of the system
is solely based on the Hamiltonian dynamics corresponding to the coarse-grained potential 
energy surface~\cite{Davtyan:2015:DFM}. Dynamical behaviour can be corrected 
by introducing additional degrees for freedom (fictitious particles)
interacting with each coarse-grained 
bead~\cite{Davtyan:2015:DFM,Erban:2016:CAM,Davtyan:2016:DFM}. 
Fictitious particles can then be subject to suitable friction and noise terms
to correct the dynamics. 

Considering our dimer molecule model as an example of a coarse-grained 
molecule, written in terms of two coarse-grained beads 
(monomers) interacting through coarse-grained potential energy~(\ref{potentialform}), 
then each monomer could be coupled with one or several fictitious particles 
interacting with the monomer through a suitable harmonic 
spring term~\cite{Davtyan:2015:DFM,Erban:2016:CAM}.
Our long-range interaction heat bath is based on this approach, 
by assuming that the $i$-th monomer, $i=1,2$, 
is coupled with $N_i$ harmonic oscillators, in 
a manner similar to well known theoretical heat bath 
models~\cite{Ford:1965:SMA,Zwanzig:1973:NGL}. Then 
equations~(\ref{nocol1})--(\ref{nocol2}),
expressing Newton's second law of motion, include additional terms 
as follows~\cite{Ford:1965:SMA}
\begin{eqnarray}
M \frac{\mbox{d}{\mathbf V}_1}{\mbox{d}t}
&=&
\Phi^\prime(R) \, \frac{{\mathbf R}}{R}
+
\sum_{j=1}^{N_1}
k_{1,j} \, \alpha_{1,j} \Big( {\mathbf x}_1^j-\alpha_{1,j} {\mathbf X}_1 \Big),
\label{dimerharmosc1}
\\
M \frac{\mbox{d}{\mathbf V}_2}{\mbox{d}t}
&=&
- \Phi^\prime(R) \, \frac{{\mathbf R}}{R}
+
\sum_{j=1}^{N_2}
k_{2,j} \, \alpha_{2,j} \Big( {\mathbf x}_2^j-\alpha_{2,j}{\mathbf X}_2 \Big),
\qquad
\label{dimerharmosc2}
\end{eqnarray}
where ${\mathbf x}_i^j$ is the position of the $j$-th solvent 
particle which interacts with the $i$-th monomer through 
a harmonic spring with spring constant $k_{i,j}$ and 
interaction constants $\alpha_{i,j}$, $j=1,2,\dots,N_i$, $i=1,2$. 
Equations~(\ref{dimerharmosc1})--(\ref{dimerharmosc2})
are coupled with the evolution equations for solvent particles. We assume 
that ${\mathbf v}_i^j$ is the velocity of the $j$-th solvent particle 
interacting with the $i$-th monomer. Moreover, we assume that all 
oscillators have the same mass, $m$. Using Newton's second law of motion, 
we get the following evolution equations for the heat bath oscillators
\begin{eqnarray}
\frac{\mbox{d}{\mathbf x}_i^j}{\mbox{d}t} 
&=& 
{\mathbf v}_i^j,
\label{solventharmosc1}
\\
m 
\frac{\mbox{d}{\mathbf v}_i^j}{\mbox{d}t}
&=&
-k_{i,j} \, \Big( {\mathbf x}_i^j -\alpha_{i,j}{\mathbf X}_i \Big),
\label{solventharmosc2}
\end{eqnarray}
for $j=1,2,\dots,N_i$ and $i=1,2.$ Unlike in some fictitious particle
models~\cite{Davtyan:2015:DFM,Erban:2016:CAM,Davtyan:2016:DFM}, we
do not include friction and random forces into equation
(\ref{solventharmosc2}) for solvent, because we assume that 
we explicitly model all solvent particles,
i.e. $N_1$ and $N_2$ are considered to
satisfy $N_1 \gg 1$ and $N_2 \gg 1$. We are therefore
working `close' to the limit $N_1 \to \infty$ and $N_2 \to \infty$,
in which we can get the convergence of our long-range interaction
heat bath to the Langevin dynamics as discussed below. In practice, it is impossible to 
include all solvent molecules in simulations and friction and noise
terms are still included to control temperature of the simulated
system~\cite{Leimkuhler:2015:MD,Leimkuhler:2009:GST}. 
We can solve the solvent equations of motion
(\ref{solventharmosc1})--(\ref{solventharmosc2}) 
to give~\cite{Leimkuhler:2015:MD,Erban:2019:SMR}
\begin{eqnarray*}
{\mathbf x}_i^j
&=&
{\mathbf x}_i^j(0) \cos \left( \omega_{i,j} t \right) + 
\frac{{\mathbf v}_i^j(0)}{\omega_{i,j}} \sin \left( \omega_{i,j} t \right)
\nonumber
\\&& \; + \; \alpha_{i,j}\, \omega_{i,j}\int_{0}^{t} 
{\sin \left( \omega_{i,j} (t-\tau) \right) {\mathbf X}_i(\tau)} \, {\mbox{d}\tau}  
\end{eqnarray*}
where ${\mathbf x}_i^j(0)$ is the initial position of the $j$-th heat bath 
particle corresponding to the $i$-th monomer, ${\mathbf v}_i^j(0)$ is its 
initial velocity and $\omega_{i,j} = (k_{i,j}/m)^{1/2}$ is its frequency. 
Substituting for ${\mathbf x}_1^j$ and ${\mathbf x}_2^j$ in dimer's equations 
of motion (\ref{dimerharmosc1})--(\ref{dimerharmosc2}), 
we obtain the following coupled system of generalized 
Langevin equations
\begin{eqnarray}
M \frac{\mbox{d}{\mathbf V}_1}{\mbox{d}t}
&=&
\Phi^\prime(R) \, \frac{{\mathbf R}}{R}
-\!\!
\int_{0}^{t} \!\!\!
\kappa_1(\tau)  \, {\mathbf V}_1(t - \tau)
\, \mbox{d}\tau
+
{\boldsymbol \xi}_1,
\label{dimergle1}
\\
M \frac{\mbox{d}{\mathbf V}_2}{\mbox{d}t}
&=&
- \Phi^\prime(R) \, \frac{{\mathbf R}}{R}
-\!\!
\int_{0}^{t} \!\!\!
\kappa_2(\tau)  \, {\mathbf V}_2(t - \tau)
\, \mbox{d}\tau
+
{\boldsymbol \xi}_2, \quad
\;\;\; \label{dimergle2}
\end{eqnarray}
where the friction kernel $\kappa_i(\tau)$ and noise term 
${\boldsymbol \xi}_i \equiv {\boldsymbol \xi}_i(t)
= [\xi_{i;1},\xi_{i;2},\xi_{i;3}]$ are given by
\begin{eqnarray*}
\kappa_i(\tau)
&=&
m
\sum_{j=1}^{N_i}
\alpha_{i,j}^2\, \omega_{i,j}^2
\cos \left( \omega_{i,j} \tau \right),
\label{gammataudef}
\\
{\boldsymbol \xi}_i(t)
& = &
m \sum_{j=1}^{N_i}
{\mathbf x}_i^j(0) \,  
\alpha_{i,j}\, \omega_{i,j}^2  
\cos \left( \omega_{i,j} t \right) 
\nonumber \\
&& \qquad \; +\;
{\mathbf v}_i^j(0) \, \alpha_{i,j}\, \omega_{i,j}  
\sin \left( \omega_{i,j} t \right), \quad
\nonumber
\end{eqnarray*}
for $i=1,2$. We assume that initial positions and velocities of solvent oscillators,
${\mathbf x}_i^j(0)$ and ${\mathbf v}_i^j(0)$, are both independently 
sampled according to their equilibrium distributions.
Then noise autocorrelation function is given by the generalized
fluctuation-dissipation theorem
\begin{equation*}
\lim_{t \to \infty}
\langle \xi_{i;j}(t) \, \xi_{i;n}(t-\tau) \rangle
=
2 k_B T
\,
\delta_{j,n}
\,
\kappa_i(\tau),
\end{equation*}
where $k_B$ is the Boltzmann constant and $T$ is the absolute 
temperature. Next, we assume that the frequencies $\omega_{i,j}$ 
are sampled from a (continuous) exponential distribution
with mean $\overline{\omega}$ and we set our interaction constants equal to
\begin{equation}
\alpha_{i,j} = \frac{1}{\omega_{i,j}} \sqrt{\frac{2 \, \gamma \, \overline{\omega}}{N_i \, m \, \pi}},
\label{alphaijequation}
\end{equation} 
where $\gamma>0$ is the friction constant used in 
equations~(\ref{langeqV1}) and (\ref{langeqV2}). 
Then friction kernel (\ref{gammataudef}) becomes
$$
\kappa_i(\tau)
=
\frac{2 \gamma \, \overline{\omega}}{\pi}
\frac{1}{N_i}
\sum_{j=1}^{N_i}
\cos \left( \omega_{i,j} \tau \right).
$$
Passing to the limit $N_i \to \infty$ allows us 
to consider the above summation as a continuous integral over the 
distribution of oscillator frequencies, with both friction 
kernels $\kappa_1(\tau)$ and $\kappa_2(\tau)$ converging to 
the same friction kernel~\cite{Erban:2019:SMR}
\begin{eqnarray}
\kappa(\tau)
&=&
\frac{2 \gamma}{\pi}
\int_{0}^\infty
\!
\cos \left( \omega \tau \right)
\, \exp\left( - \frac{\omega}{\overline{\omega}} \right) \,
\mbox{d}\omega
\nonumber
\\
&=&
\frac{2 \gamma}{\pi}
\frac{\overline{\omega}}{\overline{\omega}^2 \tau^2 + 1}.
\label{kappatau}
\end{eqnarray}
Then
$
\int_0^\infty \kappa(\tau) \, \mbox{d}\tau = \gamma.
$
Moreover, we can define the limiting friction kernel by
\begin{equation*}
\kappa_{\infty}(\tau) = \lim_{\overline{\omega} \to \infty} 
\kappa(\tau),
\end{equation*}
which, for our choice of oscillators' 
frequencies and interaction terms~(\ref{alphaijequation}),
satisfies $\kappa_{\infty}(\tau) = 0$ for $\tau>0$ and 
$\kappa_{\infty}(0) = \infty.$ Thus the limiting kernel 
is a multiple of the
Dirac delta function. Therefore the position and velocity of the 
monomers, ${\mathbf X}_i$ and ${\mathbf V}_i$, converge 
to the solution of (\ref{langeqX1})--(\ref{langeqV2}) in the 
limit $\overline{\omega} \to \infty$, provided that each monomer
has its own separate heat bath. Moreover, we obtain
the Einstein-Smoluchowski relation for the diffusion constant 
of the monomer as $D = k_B T / (\gamma \, M)$.

As in Section~\ref{sec3}, we have explained our MD model of the dimer
using the case where each monomer has its own heat bath. We now
turn our attention to the case when monomers share their heat bath.
This has been studied in the case of the short-range interaction MD model
in Section~\ref{shramulti} with the help of multi-resolution modelling
in a co-moving frame, as schematically shown in Figure~\ref{figure1}(a). 
In the case of long-range interactions, a co-moving frame is less
straightforward to implement because we need to take into account
that particles outside of the simulated box do exert (long-range)
forces on particles in our simulation domain. Some multi-resolution 
techniques in the literature solve this problem by introducing suitable 
overlap (bridging, blending) 
regions~\cite{Miller:UFP:2009,Kevrekidis:2003:EFM,Biyikli:2014:MMI,Erban:2016:CAM},
where molecules which are near the simulation domain exert some partial 
forces on the simulated molecules. 

\begin{figure}
\centerline{\hskip 5mm
\epsfig{file=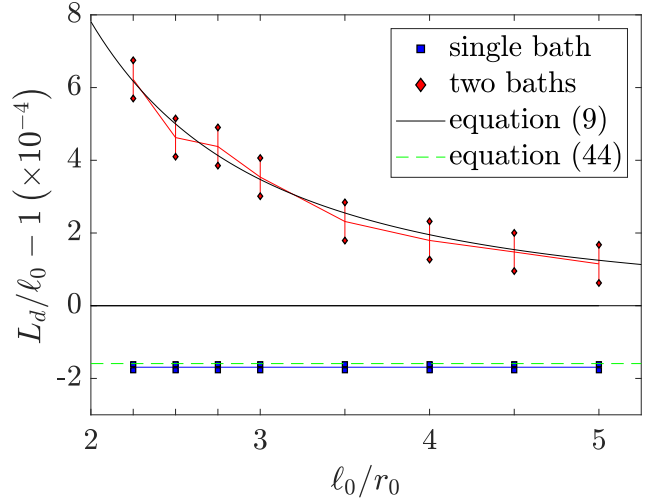,height=7cm}
}		
\caption{{\it The extension of the average length of a dimer from its 
separation distance $\ell_0$ for long-range interaction heat
bath models.  The values of parameters are the same as in Figure~$\ref{figure2}$, 
together with $\overline{\omega} = 100$, $N_1 = N_2 = N = 10^4$,
$M=1$ and $m=10^{-3}$, which give 
the same value of $\mu = M/m$ as used in Figure~$\ref{figure2}$. The simulations 
for the single heat bath case use parameter choice $(\ref{kalphasame})$ 
with $\alpha_{j}^2 = \gamma \, \overline{\omega}/(N \, \pi \, k_j)$,
$k_j = m \, \omega_j^2/2$, and $\omega_j$ sampled according to 
the exponential distribution with mean $\overline{\omega}$, confirming result 
given in equation $(\ref{singlebathhamiltonian})$ (green dashed line).
The results for the two heat bath case are compared with the result
obtained for the Langevin model in equation $(\ref{Ldasympt})$ (black solid line).
}
\label{figure5}}
\end{figure}

In what follows, we do not truncate the simulated domain, but we
consider a different multi-resolution approach in Section~\ref{sec4multi}. 
Before then we discuss results comparable to Figure~\ref{figure2}, i.e.
we compare simulations with a single heat bath and two heat baths
for the case of our long-range interaction MD model. The results are
presented in Figure~\ref{figure5}, where we use the same values of
$\ell_0$ as in Figure~\ref{figure2}, expressed as $\alpha$-multiples
of $r_0$, although our long-range interaction model does not make use
of parameter $r_0$. 
The value of $L_d$ is for each value of $\ell_0$ calculated from a long
simulation over 200 dimensionless time units, where the first 100 time 
units are used to equilibrate the system, while the second half of each 
simulation is used to compute $L_d$.
To initialise this model we start with monomers separated by the rest
length $\ell_0$ and sample oscillators' frequencies, $\omega_{i,j}$, 
according to the exponential distribution with mean $\overline{\omega}$. 
Their positions and velocities are sampled from 
the Maxwell-Boltzmann distribution. 
For the two-bath model, each dimer particle is separately 
initialised with its own set of oscillators around their 
respective positions in space. 

In Figure~\ref{figure5}, we observe that in the case of the two-bath model
we obtain results which match well with equation (\ref{Ldasympt}) for
our parameter values. These results are also directly comparable with 
the results obtained for the two-bath case in Figure~\ref{figure2}. 
The situation is more complicated in the case of simulations with
a single heat bath with $N$ oscillators. Then, using notation (\ref{singleheatbath}), 
we can rewrite (\ref{dimerharmosc1})--(\ref{dimerharmosc2}) as
\begin{equation*}
M \frac{\mbox{d}{\mathbf V}_i}{\mbox{d}t}
=
(-1)^{i+1}
\Phi^\prime(R) \, \frac{{\mathbf R}}{R}
+
\sum_{j=1}^{N}
k_{i,j} \, \alpha_{i,j} \Big( {\mathbf x}^j-\alpha_{i,j} {\mathbf X}_i \Big),
\label{dimerharmoscshb}
\end{equation*}
for $i=1,2$, where the heat bath evolution equation (\ref{solventharmosc2}) includes terms
corresponding to both monomers
\begin{equation*}
m 
\frac{\mbox{d}{\mathbf v}^j}{\mbox{d}t}
=
-
k_{1,j} \, \Big( {\mathbf x}^j -\alpha_{1,j}{\mathbf X}_1 \Big)
-
k_{2,j} \, \Big( {\mathbf x}^j -\alpha_{2,j}{\mathbf X}_2 \Big),
\end{equation*}
for $j=1,2,\dots,N$. Our results will then depend how we choose
parameters $k_{i,j}$ and $\alpha_{i,j}$. For example, if we
choose $k_{i,j}$ and $\alpha_{i,j}$ to be the same for
both monomeres, i.e.
\begin{equation}
k_{1,j} = k_{2,j} = k_{j}, \quad \mbox{and} \quad
\alpha_{1,j} = \alpha_{2,j} = \alpha_j,
\label{kalphasame}
\end{equation}
for $j=1,2,\dots,N,$ then the oscillating frequency of the $j$-th heat bath oscillator
is $\omega_{j} = \sqrt{2 k_{j}/m}$ and we can subtract the evolution equations
for monomers to obtain
$$
M 
\frac{\mbox{d}^2{\mathbf R}}{\mbox{d}t^2}
=
- 2
\Phi^\prime(R) \, \frac{{\mathbf R}}{R}
-
\sum_{j=1}^{N}
k_{j} \, \alpha_{j}^2 \, {\mathbf R}.
$$
This equation does not contain any heat bath variables. Using 
(\ref{alphaijequation}) to select $\alpha_{j}$, i.e. using 
$\alpha_{j}^2 
= 
2 \, \gamma \, \overline{\omega}/(N \, m\, \pi \, \omega_j^2)
=
\gamma \, \overline{\omega}/(N \, \pi \, k_j)$,
we get
$$
M 
\frac{\mbox{d}^2{\mathbf R}}{\mbox{d}t^2}
=
- 2
\Phi^\prime(R) \, \frac{{\mathbf R}}{R}
+
\frac{\gamma \, \overline{\omega}}{\pi} \, {\mathbf R}.
$$
Using potential (\ref{potentialform}), we conclude that we effectively
obtain a shorter rest length of the spring which gives the following
approximation
\begin{equation}
L_d \,\approx\, \frac{2 \, k \, \pi \, \ell_0 }{2\, k \,\pi + \gamma \, \overline{\omega}}.
\label{singlebathhamiltonian}
\end{equation}
This result is plotted in Figure~\ref{figure5} together with
results obtained by illustrative simulations. We use a long-time 
simulation of length $200$ dimensionless time units, with monomers 
initially placed at separation $\ell_0$ and averaging over the 
second half of the simulation (of length $100$ dimensionless 
time units) to obtain the presented values of dimer's
expected length $L_d$.

In Figure~\ref{figure5}, we observe that the average dimer length, 
$L_d$, during our single heat bath simulations is smaller than the 
natural length of the spring, $\ell_0$. However, this conclusion is 
only a consequence of our choice of parameters (\ref{kalphasame}). 
An opposite phenomenon can be observed in simulations for other 
parameter regimes. For example, if we divide our oscillators into two groups 
consisting of $N_1$ and $N_2$ oscillators, i.e. $N=N_1+N_2$, and 
choose our parameters $k_{i,j}$ and $\alpha_{i,j}$ such that
\begin{eqnarray*}
&k_{2,j} = 0, &\qquad \mbox{for} \; j=1,2,\dots,N_1, \\
&k_{1,j} = 0, &\qquad \mbox{for} \; j=N_1+1,N_1+2,\dots,N,
\end{eqnarray*}
then our ``one-bath" case is effectively equal to the two-bath 
case for which we have the result given in equation (\ref{Ldasympt})
presented in Figure~\ref{figure5}. In particular, depending on our 
choices of $k_{i,j}$ and $\alpha_{i,j}$, the single heat bath case 
can both increase or decrease the average length of the dimer.

\subsection{Multi-resolution modelling of dimer}

\label{sec4multi}

\begin{table}

\framebox{%
\hsize=0.961\hsize
\vbox{
\leftskip 7.6mm
\parindent -7.6mm

[L1] \hskip 0.6mm
Update velocities of the dimer and solvent particles for a half time step using (\ref{velverlet1}).
			
\smallskip
			
[L2] \hskip 0.6mm
Update positions of the dimer and solvent particles using (\ref{velverlet2}).
			
\smallskip
			
[L3] \hskip 0.6mm
Recalculate accelerations of each monomer and solvent oscillators by (\ref{defA1}),
(\ref{langeqA2disc}) and (\ref{defaj}).
			
\smallskip
			
[L4] \hskip 0.6mm
Update velocities of the dimer and solvent particles for a half time step using (\ref{velverlet3}).
			
\smallskip
			
[L5] \hskip 0.6mm
Continue with step [L1] using time $t=t+\Delta t$.

\par \vskip 0.8mm}
} \caption{\label{table4} 
One iteration of the multi-resolution simulation algorithm of 
the dimer in the heat bath with long-range interactions, 
where the second monomer is simulated by the Langevin dynamics.}
\end{table}

\noindent
In Figure~\ref{figure1}(b), we use our dimer example to illustrate a multi-resolution 
approach which models a part of a molecule using a detailed MD approach, while using 
a coarser description of the rest of the molecule. Here, 
in the same manner as carried out for our short-range 
model in Section~\ref{secmondifres}, we illustrate such a
multi-resolution approach using our long-range interaction MD model. We use the 
Langevin model~(\ref{langeqX1})--(\ref{langeqV2}) to coarse-grain one of the monomers, 
while the other monomer is modelled in detail using the MD model with its heat bath
described by harmonic oscillators (\ref{solventharmosc1})--(\ref{solventharmosc2}). 
As in Figure~\ref{figure3}, we again calculate numerical estimates for the velocity 
autocorrelation function, $C_d(\tau)$ in equation (\ref{vacfdef}), from long time 
simulations of the dimer after equilibrium has been reached. 

\begin{figure}
\centerline{\hskip 5mm
\epsfig{file=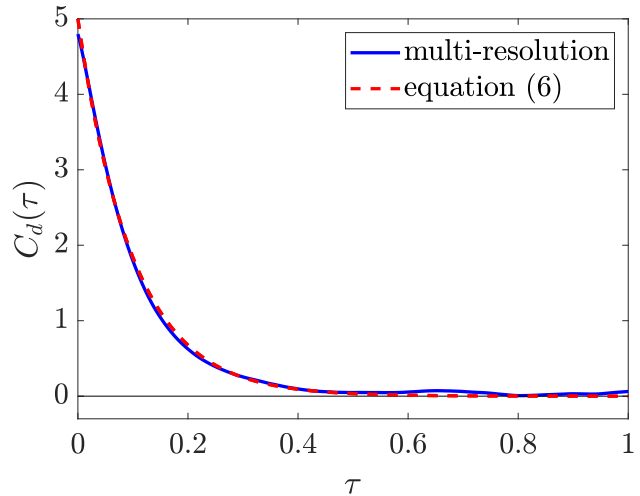,height=7cm}
}		
\caption{{\it 
The velocity autocorrelation function 
for the multi-resolution model (blue solid line) 
for long-range interactions, estimated 
from long time simulation over dimensionless 
time of $10^3$ dimensionless time units. It is compared with the result
for the Langevin description of the whole dimer, given
by equation $(\ref{Cdform})$ (red dashed line).
The parameters are the same as in Figure~$\ref{figure3}$,
namely $\gamma=10$, $D=1$, $k=10^{6}$, $M = 1$, $m=10^{-3}$,
$\ell_0 = 0.32$, together with $\overline{\omega} = 100$ 
and $N_1 = 10^5$.}
\label{figure6}}
\end{figure}

The pseudo-code of one iteration our multi-resolution algorithm is presented 
as Algorithm~[L1]--[L5] in Table~\ref{table4}. Algorithm~[L1]--[L5] is based 
on the velocity Verlet integrator, where both monomers are updated by
\begin{eqnarray}
{\mathbf V}_i\left(t+\tfrac{1}{2}\Delta t\right) 
&=& 
{\mathbf V}_i(t) + \frac{1}{2} \, {\mathbf A}_i(t) \, \Delta t,
\label{velverlet1}
\\
{\mathbf X}_i(t+\Delta t) 
&=& 
{\mathbf X}_i(t) +  {\mathbf V}_i(t+\tfrac{1}{2}\Delta t) \, \Delta t, 
\label{velverlet2}
\\	
{\mathbf V}_i(t+\Delta t) 
&=& 
{\mathbf V}_i\left(t+\tfrac{1}{2}\Delta t\right) + \frac{1}{2} \, {\mathbf A}_i(t + \Delta t) \, \Delta t,
\qquad\;\;
\label{velverlet3}
\end{eqnarray}
where ${\mathbf A}_i$, for $i=1,2$, is the acceleration of the corresponding
monomer. For the first monomer, its acceleration ${\mathbf A}_1$ is
defined as the right hand side of equation~(\ref{dimerharmosc1}) 
divided by $M$, i.e.
\begin{equation}
{\mathbf A}_1
=
\frac{\Phi^\prime(R)}{M} \, \frac{{\mathbf R}}{R}
+
\frac{1}{M}
\sum_{j=1}^{N_1}
k_{1,j} \, \alpha_{1,j} \Big( {\mathbf x}_1^j-\alpha_{1,j} {\mathbf X}_1 \Big).
\label{defA1}
\end{equation}
For the second monomer, we use the BBK integrator~\cite{Brunger:1984:SBC},
i.e. we define its acceleration as
\begin{equation}
{\mathbf A}_2
=
- 
\frac{\Phi^\prime(R)}{M} 
\, \frac{{\mathbf R}}{R}
-
\gamma {\mathbf V}_2    
+ 
\gamma \, \sqrt{\frac{2 D }{\Delta t}} \; \xi,
\label{langeqA2disc}
\end{equation}
where $\xi$ is sampled from the normal distribution with zero mean and unit
variance. The corresponding solvent oscillator integrator is identical
to the scheme (\ref{velverlet1})--(\ref{velverlet3}), 
with ${\mathbf X}_1$, ${\mathbf V}_1$ and ${\mathbf A}_1$ replaced by 
${\mathbf x}^j$, ${\mathbf v}^j$ and ${\mathbf a}^j$, respectively, where
acceleration ${\mathbf a}^j$ is defined as the right hand side of 
equation~(\ref{solventharmosc2}) divided by $m$, i.e.
\begin{equation}
{\mathbf a}^j
=
- 
\frac{k_{i,j}}{m} 
\, \Big( {\mathbf x}_i^j -\alpha_{i,j}{\mathbf X}_i \Big).
\label{defaj}
\end{equation}
The results obtained by Algorithm~[L1]--[L5] are compared with analytic 
results given by equation (\ref{Cdform}) for the Langevin model 
in Figure~\ref{figure6}. We see that there is a good correspondence 
between these, suggesting that the value $\bar{\omega}=100$ is large 
enough to create an accurate Dirac delta approximation from the 
kernel function~(\ref{kappatau}), along with having a large enough 
number of oscillators, $N_1 = 10^5$, in 
our heat bath for our other approximations to hold. If these conditions 
did not hold, we would see that our kernel function has a different 
form (for example, decaying at a slower rate), and in this case 
we would have to use a generalized Langevin model as our 
coarse-graining approach in order to capture the dynamics 
of the dimer with sufficient accuracy. 

The diffusion constant of the dimer, $D_d$, can again be estimated by 
numerically integrating the velocity auto-correlation function.
Integrating our results from Figure~\ref{figure6} over interval $[0,1]$, 
we obtain $D_d \approx 0.510$, which compares well with the theoretical 
value, $D/2 = 0.5$, given by equation (\ref{Ddvalue}).

\section{Discussion and Conclusions}

\label{sec5}

\noindent
In this paper, we have used two theoretical heat baths.
Although these heat baths are based on qualitatively different descriptions of 
solvent-dimer interactions, they both lead to the Langevin description, given in 
equations~(\ref{langeqX1})--(\ref{langeqV2}), in a certain limit. In particular,
we can use this limiting process to coarse-grain a part of the simulated dimer
molecule, while use a detailed MD model to describe the rest of the molecule.
Such a multi-resolution approach has potential to significantly speed up
computer simulations of dynamics of 
macromolecules~\cite{Fogarty:2016:MMC,Rolls:2017:VRR,Rolls:2018:MPB,DiPasquale:2014:MTS,DiPasquale:2017:LGD},
provided that it is combined with additional multiscale and multi-resolution 
methodologies, discussed below.

Our long-range interaction model leads to the system of generalized Langevin 
equations, given by equations~(\ref{dimergle1})--(\ref{dimergle2}). Although 
we have worked in the parameter regime where the generalized Langevin 
equations can be well approximated by the system of Langevin equations 
given by~(\ref{langeqX1})--(\ref{langeqV2}), this will not be the case in 
other parameter regimes and for more realistic solvent descriptions, 
especially when the memory kernel is estimated from MD 
simulations~\cite{Jung:2017:IRM,Shin:2010:BMM}. One possible strategy
in this case is to couple a detailed MD model with a stochastic 
coarse-grained model which is written with the help of additional 
variables~\cite{Davtyan:2015:DFM,Erban:2016:CAM,Davtyan:2016:DFM}.
To improve the efficiency of simulations further, one can then coarse-grain
such a generalized Langevin description using a Brownian
dynamics approach~\cite{Erban:2009:SMR,Erban:2014:MDB}. Brownian dynamics
modelling can be further coupled with stochastic reaction-diffusion
modelling based on lattice-based (compartment-based) methods~\cite{Robinson:2015:MRS}.
Lattice-based models are very attractive for simulations of intracellular
processes, because they enable modelling of spatio-temporal processes in
the whole cell or its significant part~\cite{Roberts:2013:LMH}. Coupling
Brownian dynamics with compartment-based approaches has been used in 
a number of applications, including multi-resolution
modelling of actin dynamics in filopodia~\cite{Erban:2014:MSR,Dobramysl:2016:SEI}
or for modelling intracellular calcium dynamics~\cite{Dobramysl:2016:PMM}.

In this paper, we have investigated multi-resolution approaches, 
schematically described in Figure~\ref{figure1}(a) and 1(b). Another class 
of multi-resolution approaches in the literature considers a fixed subdomain 
of the computational domain where a detailed modelling approach is used, which 
is coupled with a coarser model in the rest of the simulation 
domain~\cite{Flegg:2012:TRM,Robinson:2015:MRS}. Such an approach is useful,
for example, when modelling intracellular ion dynamics. 
Ions pass through an ion channel in single file and an MD model has to be used 
to accurately compute the discrete, stochastic, current in the 
channel~\cite{Allen:2000:MDE,Jensen:2010:PCH}, while the details of the behaviour 
of individual  ions are less important away from the channel where copy numbers 
may be very large. Thus, we can improve efficiency of our simulations if we allow 
ions to pass between regions with an explicitly modelled heat bath and a region 
where their trajectories are described by coarser stochastic 
models~\cite{Erban:2016:CAM}. 

A similar multi-resolution approach can 
also be designed for our illustrative dimer model. It is schematically
shown in Figure~\ref{figure1}(c), where we identify the region with explicitly 
simulated heat bath as $\{x_1 > b\} = (b,\infty) \times {\mathbb R}^2$, 
where $b$ is the fixed position of the boundary.
We are again interested in the behaviour of the dimer in the MD 
model which would be considered in the full space, ${\mathbb R}^3$. 
However, we now want to replace solvent particles which are 
in $\{x_1 < b\} = (-\infty,b) \times {\mathbb R}^2$ by a coarser Langevin description~\eqref{langeqX1}--\eqref{langeqV2}. 
To do that, we have to carefully consider how we handle the
transfer of monomers between $\{x_1 > b\}$ and $\{x_1 < b\}$.
In Figure~\ref{figure1}(c), we present a two-dimensional illustration
of a monomer when it intersects the interface, $\{x_1 = b\}$. 
Such a monomer is subject to the collisions with heat bath particles 
on the part of its surface which lies in $\{x_1 > b\}$. This has to 
be compensated by using a suitable random force from 
$\{x_1 < b\}$, so that the overall 
model is equivalent to~\eqref{langeqX1}--\eqref{langeqV2}
in the Langevin limit. Such correction terms can be derived analytically
for the case of a spherical monomer in our short-range interaction
heat bath and are presented in 
References~\cite{Erban:2014:MDB,Erban:2019:SMR}. They can be used to couple
the MD model with its corresponding Langevin description, which can be further
coupled with Brownian dynamics, simulated using a much larger
time step~\cite{Erban:2014:MDB}.

Mathematical analysis of multi-resolution methodologies can make use 
of the analysis of the model behaviour close to the boundaries of the 
computational domain. For example, derivations of reactive (Robin) 
boundary conditions of macroscopic models from their corresponding 
microscopic descriptions~\cite{Erban:2007:RBC,Erban:2007:TSR,Chapman:2016:RBC} 
can be generalized to the analysis of behaviour of molecules close
to hybrid interfaces in multi-resolution 
schemes~\cite{Flegg:2012:TRM,Flegg:2014:ATM,Flegg:2015:CMC}. Analysis
of open boundaries of MD schemes (i.e. boundaries which can transfer
mass, momentum and energy) can lead to further understanding of multi-resolution
schemes such as AdResS and hybrid continuum-particle dynamics~\cite{Delgado:2015:OBM},
which enable efficient simulation of biomolecules at realistic physiological 
conditions~\cite{Zavadlav:2018:ARS}.

Equations for coupled detailed/coarse-grained models can be systematically
derived using Zwanzig's projection method, which has been used to address
co-existence of atoms and beads (larger coarse-grained units) in the same
dynamic simulations~\cite{Espanol:2009:HDC,DiPasquale:2018:SDH}. The equations
of motion take the form of dissipative particle dynamics, which have
been coupled with atomistic water simulations to design multi-resolution 
schemes in the literature~\cite{Zavadlavb:2017:ARS}. Other multi-resolution 
methods couple atomistic water with specially designed coarse-grained water
models~\cite{Gonzalez:2013:TMA} or with a continuum 
approach~\cite{Delgado:2009:CAC}. Coupling discrete and continuum approaches
can also be done for different molecular species present in the system
and our choice of a modelling approach for each species can be based on its
relative abundance~\cite{Liu:2012:HMS,Duncan:2016:HFS,Franz:2012:HMI}.

One of several important points which have been left out from our discussion 
is the discretization of time. Although our illustrative simulations use
the same time step for both the MD model and the Langevin description, this
is not the most efficient or desirable strategy, because the MD model requires 
much smaller time step than the corresponding Langevin equation. There is
potential to design more efficient schemes by updating the coarser
description only at certain multiples of the time step which is used in 
the most detailed model~\cite{Rolls:2017:VRR}. This is also the
case when a modeller further coarse-grains the Langevin description into a 
Brownian dynamics model which uses even large timesteps~\cite{Erban:2014:MDB}. 

\vskip -8mm
\rule{0pt}{0pt}

\subsection*{Acknowledgements}

\vskip -4mm

\noindent
This work was supported by funding from the Engineering and Physical Sciences 
Research Council (EPSRC) [grant number EP/G03706X/1]. Radek Erban would also
like to thank the Royal Society for a University Research Fellowship.

\vskip -8mm
\rule{0pt}{0pt}

\subsection*{Authors' contributions}

\vskip -4mm

\noindent
Ravinda Gunaratne (RG) and Daniel Wilson (DW) wrote computer codes and peformed 
computer simulations to obtain illustrative results presented in Figure~\ref{figure2} (DW, RG), 
Figure~\ref{figure3} (DW, RG), Figure~\ref{figure5} (RG) and Figure~\ref{figure6} (RG). 
All authors worked on the analysis of both (short-range and long-range 
interaction) MD models, wrote the paper and gave final approval for publication.

\vskip -8mm
\rule{0pt}{0pt}

\subsection*{Data accessibility}

\vskip -4mm

\noindent
The computer codes used to compute illustrative results in 
Figures~\ref{figure2}, \ref{figure3}, \ref{figure5} and \ref{figure6} 
have been uploaded as part of the supplementary
material. Figures~\ref{figure1} and \ref{figaccprob} contain no data.

\section*{References}

\rule{0pt}{1pt}

\vskip -1.5cm

\rule{0pt}{1pt}


\begin{thebibliography}{78}%
\makeatletter
\providecommand \@ifxundefined [1]{%
 \@ifx{#1\undefined}
}%
\providecommand \@ifnum [1]{%
 \ifnum #1\expandafter \@firstoftwo
 \else \expandafter \@secondoftwo
 \fi
}%
\providecommand \@ifx [1]{%
 \ifx #1\expandafter \@firstoftwo
 \else \expandafter \@secondoftwo
 \fi
}%
\providecommand \natexlab [1]{#1}%
\providecommand \enquote  [1]{``#1''}%
\providecommand \bibnamefont  [1]{#1}%
\providecommand \bibfnamefont [1]{#1}%
\providecommand \citenamefont [1]{#1}%
\providecommand \href@noop [0]{\@secondoftwo}%
\providecommand \href [0]{\begingroup \@sanitize@url \@href}%
\providecommand \@href[1]{\@@startlink{#1}\@@href}%
\providecommand \@@href[1]{\endgroup#1\@@endlink}%
\providecommand \@sanitize@url [0]{\catcode `\\12\catcode `\$12\catcode
  `\&12\catcode `\#12\catcode `\^12\catcode `\_12\catcode `\%12\relax}%
\providecommand \@@startlink[1]{}%
\providecommand \@@endlink[0]{}%
\providecommand \url  [0]{\begingroup\@sanitize@url \@url }%
\providecommand \@url [1]{\endgroup\@href {#1}{\urlprefix }}%
\providecommand \urlprefix  [0]{URL }%
\providecommand \Eprint [0]{\href }%
\providecommand \doibase [0]{http://dx.doi.org/}%
\providecommand \selectlanguage [0]{\@gobble}%
\providecommand \bibinfo  [0]{\@secondoftwo}%
\providecommand \bibfield  [0]{\@secondoftwo}%
\providecommand \translation [1]{[#1]}%
\providecommand \BibitemOpen [0]{}%
\providecommand \bibitemStop [0]{}%
\providecommand \bibitemNoStop [0]{.\EOS\space}%
\providecommand \EOS [0]{\spacefactor3000\relax}%
\providecommand \BibitemShut  [1]{\csname bibitem#1\endcsname}%
\let\auto@bib@innerbib\@empty
\bibitem [{\citenamefont {Rapaport}(2004)}]{Rapaport:2004:AMD}%
  \BibitemOpen
  \bibfield  {author} {\bibinfo {author} {\bibfnamefont {D.}~\bibnamefont
  {Rapaport}},\ }\href@noop {} {\emph {\bibinfo {title} {The Art of Molecular
  Dynamics Simulation}}}\ (\bibinfo  {publisher} {Cambridge University Press},\
  \bibinfo {year} {2004})\BibitemShut {NoStop}%
\bibitem [{\citenamefont {Leimkuhler}\ and\ \citenamefont
  {Matthews}(2015)}]{Leimkuhler:2015:MD}%
  \BibitemOpen
  \bibfield  {author} {\bibinfo {author} {\bibfnamefont {B.}~\bibnamefont
  {Leimkuhler}}\ and\ \bibinfo {author} {\bibfnamefont {C.}~\bibnamefont
  {Matthews}},\ }\href@noop {} {\emph {\bibinfo {title} {Molecular
  {D}ynamics}}},\ \bibinfo {series} {Interdisciplinary Applied Mathematics},
  Vol.~\bibinfo {volume} {39}\ (\bibinfo  {publisher} {Springer},\ \bibinfo
  {year} {2015})\BibitemShut {NoStop}%
\bibitem [{\citenamefont {Marrink}\ \emph {et~al.}(2007)\citenamefont
  {Marrink}, \citenamefont {Risselada}, \citenamefont {Yefimov}, \citenamefont
  {Tieleman},\ and\ \citenamefont {de~Vries}}]{Marrink:2007:MFF}%
  \BibitemOpen
  \bibfield  {author} {\bibinfo {author} {\bibfnamefont {S.}~\bibnamefont
  {Marrink}}, \bibinfo {author} {\bibfnamefont {H.}~\bibnamefont {Risselada}},
  \bibinfo {author} {\bibfnamefont {S.}~\bibnamefont {Yefimov}}, \bibinfo
  {author} {\bibfnamefont {D.}~\bibnamefont {Tieleman}}, \ and\ \bibinfo
  {author} {\bibfnamefont {A.}~\bibnamefont {de~Vries}},\ }\href@noop {}
  {\bibfield  {journal} {\bibinfo  {journal} {Journal of Physical Chemistry B}\
  }\textbf {\bibinfo {volume} {111}},\ \bibinfo {pages} {7812} (\bibinfo {year}
  {2007})}\BibitemShut {NoStop}%
\bibitem [{\citenamefont {Yesylevskyy}\ \emph {et~al.}(2010)\citenamefont
  {Yesylevskyy}, \citenamefont {Sch\"afer}, \citenamefont {Sengupta},\ and\
  \citenamefont {Marrink}}]{Yeselevskyy:2010:PWM}%
  \BibitemOpen
  \bibfield  {author} {\bibinfo {author} {\bibfnamefont {S.}~\bibnamefont
  {Yesylevskyy}}, \bibinfo {author} {\bibfnamefont {L.}~\bibnamefont
  {Sch\"afer}}, \bibinfo {author} {\bibfnamefont {D.}~\bibnamefont {Sengupta}},
  \ and\ \bibinfo {author} {\bibfnamefont {S.}~\bibnamefont {Marrink}},\
  }\href@noop {} {\bibfield  {journal} {\bibinfo  {journal} {PLoS Computational
  Biology}\ }\textbf {\bibinfo {volume} {6}},\ \bibinfo {pages} {e1000810}
  (\bibinfo {year} {2010})}\BibitemShut {NoStop}%
\bibitem [{\citenamefont {Riniker}\ and\ \citenamefont {van
  Gunsteren}(2011)}]{Riniker:2011:SEP}%
  \BibitemOpen
  \bibfield  {author} {\bibinfo {author} {\bibfnamefont {S.}~\bibnamefont
  {Riniker}}\ and\ \bibinfo {author} {\bibfnamefont {W.}~\bibnamefont {van
  Gunsteren}},\ }\href@noop {} {\bibfield  {journal} {\bibinfo  {journal}
  {Journal of Chemical Physics}\ }\textbf {\bibinfo {volume} {134}},\ \bibinfo
  {pages} {084110} (\bibinfo {year} {2011})}\BibitemShut {NoStop}%
\bibitem [{\citenamefont {Darr\'e}\ \emph {et~al.}(2010)\citenamefont
  {Darr\'e}, \citenamefont {Machado}, \citenamefont {Dans}, \citenamefont
  {Herrera},\ and\ \citenamefont {Pantano}}]{Darre:2010:ACG}%
  \BibitemOpen
  \bibfield  {author} {\bibinfo {author} {\bibfnamefont {L.}~\bibnamefont
  {Darr\'e}}, \bibinfo {author} {\bibfnamefont {M.}~\bibnamefont {Machado}},
  \bibinfo {author} {\bibfnamefont {P.}~\bibnamefont {Dans}}, \bibinfo {author}
  {\bibfnamefont {F.}~\bibnamefont {Herrera}}, \ and\ \bibinfo {author}
  {\bibfnamefont {S.}~\bibnamefont {Pantano}},\ }\href@noop {} {\bibfield
  {journal} {\bibinfo  {journal} {Journal of Chemical Theory and Computation}\
  }\textbf {\bibinfo {volume} {6}},\ \bibinfo {pages} {3793} (\bibinfo {year}
  {2010})}\BibitemShut {NoStop}%
\bibitem [{\citenamefont {Israelachvili}(2011)}]{Israelachvili:2011:ISF}%
  \BibitemOpen
  \bibfield  {author} {\bibinfo {author} {\bibfnamefont {J.}~\bibnamefont
  {Israelachvili}},\ }\href@noop {} {\emph {\bibinfo {title} {{I}ntermolecular
  and {S}urface {F}orces}}},\ \bibinfo {edition} {3rd}\ ed.\ (\bibinfo
  {publisher} {Academic Press, Elsevier},\ \bibinfo {year} {2011})\BibitemShut
  {NoStop}%
\bibitem [{\citenamefont {Rowlinson}(2002)}]{Rowlinson:2002:CSH}%
  \BibitemOpen
  \bibfield  {author} {\bibinfo {author} {\bibfnamefont {J.}~\bibnamefont
  {Rowlinson}},\ }\href@noop {} {\emph {\bibinfo {title} {{C}ohesion: A
  {S}cientific {H}istory of {I}ntermolecular {F}orces}}}\ (\bibinfo
  {publisher} {Cambridge University Press},\ \bibinfo {year}
  {2002})\BibitemShut {NoStop}%
\bibitem [{\citenamefont {Lennard-Jones}(1924)}]{LennardJones:1924:DMF}%
  \BibitemOpen
  \bibfield  {author} {\bibinfo {author} {\bibfnamefont {J.}~\bibnamefont
  {Lennard-Jones}},\ }\href@noop {} {\bibfield  {journal} {\bibinfo  {journal}
  {Proceedings of the Royal Society of London Series A}\ }\textbf {\bibinfo
  {volume} {106}},\ \bibinfo {pages} {463} (\bibinfo {year}
  {1924})}\BibitemShut {NoStop}%
\bibitem [{\citenamefont {Lennard-Jones}(1931)}]{LennardJones:1931:C}%
  \BibitemOpen
  \bibfield  {author} {\bibinfo {author} {\bibfnamefont {J.}~\bibnamefont
  {Lennard-Jones}},\ }\href@noop {} {\bibfield  {journal} {\bibinfo  {journal}
  {Proceedings of the Physical Society}\ }\textbf {\bibinfo {volume} {43}},\
  \bibinfo {pages} {461} (\bibinfo {year} {1931})}\BibitemShut {NoStop}%
\bibitem [{\citenamefont {Holley}(1971)}]{Holley:1971:MHP}%
  \BibitemOpen
  \bibfield  {author} {\bibinfo {author} {\bibfnamefont {R.}~\bibnamefont
  {Holley}},\ }\href@noop {} {\bibfield  {journal} {\bibinfo  {journal}
  {Zeitschrift f\"ur Wahrscheinlichkeitstheorie und verwandte Gebiete}\ }\textbf
  {\bibinfo {volume} {17}},\ \bibinfo {pages} {181} (\bibinfo {year}
  {1971})}\BibitemShut {NoStop}%
\bibitem [{\citenamefont {D\"urr}, \citenamefont {Goldstein},\ and\
  \citenamefont {Lebowitz}(1981)}]{Durr:1981:MMB}%
  \BibitemOpen
  \bibfield  {author} {\bibinfo {author} {\bibfnamefont {D.}~\bibnamefont
  {D\"urr}}, \bibinfo {author} {\bibfnamefont {S.}~\bibnamefont {Goldstein}}, \
  and\ \bibinfo {author} {\bibfnamefont {J.}~\bibnamefont {Lebowitz}},\
  }\href@noop {} {\bibfield  {journal} {\bibinfo  {journal} {Communications in
  Mathematical Physics}\ }\textbf {\bibinfo {volume} {78}},\ \bibinfo {pages}
  {507} (\bibinfo {year} {1981})}\BibitemShut {NoStop}%
\bibitem [{\citenamefont {Dunkel}\ and\ \citenamefont
  {H\"anggi}(2006)}]{Dunkel:2006:RBM}%
  \BibitemOpen
  \bibfield  {author} {\bibinfo {author} {\bibfnamefont {J.}~\bibnamefont
  {Dunkel}}\ and\ \bibinfo {author} {\bibfnamefont {P.}~\bibnamefont
  {H\"anggi}},\ }\href@noop {} {\bibfield  {journal} {\bibinfo  {journal}
  {Physical Review E}\ }\textbf {\bibinfo {volume} {74}},\ \bibinfo {pages}
  {051106} (\bibinfo {year} {2006})}\BibitemShut {NoStop}%
\bibitem [{\citenamefont {Erban}(2014)}]{Erban:2014:MDB}%
  \BibitemOpen
  \bibfield  {author} {\bibinfo {author} {\bibfnamefont {R.}~\bibnamefont
  {Erban}},\ }\href@noop {} {\bibfield  {journal} {\bibinfo  {journal}
  {Proceedings of the Royal Society A}\ }\textbf {\bibinfo {volume} {470}},\
  \bibinfo {pages} {20140036} (\bibinfo {year} {2014})}\BibitemShut {NoStop}%
\bibitem [{\citenamefont {Ford}, \citenamefont {Kac},\ and\ \citenamefont
  {Mazur}(1965)}]{Ford:1965:SMA}%
  \BibitemOpen
  \bibfield  {author} {\bibinfo {author} {\bibfnamefont {G.}~\bibnamefont
  {Ford}}, \bibinfo {author} {\bibfnamefont {M.}~\bibnamefont {Kac}}, \ and\
  \bibinfo {author} {\bibfnamefont {P.}~\bibnamefont {Mazur}},\ }\href@noop {}
  {\bibfield  {journal} {\bibinfo  {journal} {Journal of Mathematical Physics}\
  }\textbf {\bibinfo {volume} {6}},\ \bibinfo {pages} {504} (\bibinfo {year}
  {1965})}\BibitemShut {NoStop}%
\bibitem [{\citenamefont {Zwanzig}(1973)}]{Zwanzig:1973:NGL}%
  \BibitemOpen
  \bibfield  {author} {\bibinfo {author} {\bibfnamefont {R.}~\bibnamefont
  {Zwanzig}},\ }\href@noop {} {\bibfield  {journal} {\bibinfo  {journal}
  {Journal of Statistical Physics}\ }\textbf {\bibinfo {volume} {9}},\ \bibinfo
  {pages} {215} (\bibinfo {year} {1973})}\BibitemShut {NoStop}%
\bibitem [{\citenamefont {Bussi}\ and\ \citenamefont
  {Parrinello}(2007)}]{Bussi:2007:ASU}%
  \BibitemOpen
  \bibfield  {author} {\bibinfo {author} {\bibfnamefont {G.}~\bibnamefont
  {Bussi}}\ and\ \bibinfo {author} {\bibfnamefont {M.}~\bibnamefont
  {Parrinello}},\ }\href@noop {} {\bibfield  {journal} {\bibinfo  {journal}
  {Physical Review E}\ }\textbf {\bibinfo {volume} {75}},\ \bibinfo {pages}
  {056707} (\bibinfo {year} {2007})}\BibitemShut {NoStop}%
\bibitem [{\citenamefont {Frenkel}\ and\ \citenamefont
  {Smit}(2002)}]{Frenkel:2002:UMS}%
  \BibitemOpen
  \bibfield  {author} {\bibinfo {author} {\bibfnamefont {D.}~\bibnamefont
  {Frenkel}}\ and\ \bibinfo {author} {\bibfnamefont {B.}~\bibnamefont {Smit}},\
  }\href@noop {} {\emph {\bibinfo {title} {Understanding {M}olecular
  {S}imulation, From {A}lgorithms to {A}pplications}}},\ \bibinfo {edition}
  {2nd}\ ed.\ (\bibinfo  {publisher} {Academic Press, Elsevier},\ \bibinfo
  {year} {2002})\BibitemShut {NoStop}%
\bibitem [{\citenamefont {Huggins}(2012)}]{Huggins:2012:CLW}%
  \BibitemOpen
  \bibfield  {author} {\bibinfo {author} {\bibfnamefont {D.}~\bibnamefont
  {Huggins}},\ }\href@noop {} {\bibfield  {journal} {\bibinfo  {journal}
  {Journal of Chemical Physics}\ }\textbf {\bibinfo {volume} {136}},\ \bibinfo
  {pages} {064518} (\bibinfo {year} {2012})}\BibitemShut {NoStop}%
\bibitem [{\citenamefont {Mark}\ and\ \citenamefont
  {Nilsson}(2001)}]{Mark:2001:SDT}%
  \BibitemOpen
  \bibfield  {author} {\bibinfo {author} {\bibfnamefont {P.}~\bibnamefont
  {Mark}}\ and\ \bibinfo {author} {\bibfnamefont {L.}~\bibnamefont {Nilsson}},\
  }\href@noop {} {\bibfield  {journal} {\bibinfo  {journal} {Journal of
  Physical Chemistry A}\ }\textbf {\bibinfo {volume} {105}},\ \bibinfo {pages}
  {9954} (\bibinfo {year} {2001})}\BibitemShut {NoStop}%
\bibitem [{\citenamefont {Flegg}, \citenamefont {Chapman},\ and\ \citenamefont
  {Erban}(2012)}]{Flegg:2012:TRM}%
  \BibitemOpen
  \bibfield  {author} {\bibinfo {author} {\bibfnamefont {M.}~\bibnamefont
  {Flegg}}, \bibinfo {author} {\bibfnamefont {J.}~\bibnamefont {Chapman}}, \
  and\ \bibinfo {author} {\bibfnamefont {R.}~\bibnamefont {Erban}},\
  }\href@noop {} {\bibfield  {journal} {\bibinfo  {journal} {Journal of the
  Royal Society Interface}\ }\textbf {\bibinfo {volume} {9}},\ \bibinfo {pages}
  {859} (\bibinfo {year} {2012})}\BibitemShut {NoStop}%
\bibitem [{\citenamefont {Robinson}, \citenamefont {Andrews},\ and\
  \citenamefont {Erban}(2015)}]{Robinson:2015:MRS}%
  \BibitemOpen
  \bibfield  {author} {\bibinfo {author} {\bibfnamefont {M.}~\bibnamefont
  {Robinson}}, \bibinfo {author} {\bibfnamefont {S.}~\bibnamefont {Andrews}}, \
  and\ \bibinfo {author} {\bibfnamefont {R.}~\bibnamefont {Erban}},\
  }\href@noop {} {\bibfield  {journal} {\bibinfo  {journal} {Bioinformatics}\
  }\textbf {\bibinfo {volume} {31}},\ \bibinfo {pages} {2406} (\bibinfo {year}
  {2015})}\BibitemShut {NoStop}%
\bibitem [{\citenamefont {Praprotnik}\ \emph {et~al.}(2007)\citenamefont
  {Praprotnik}, \citenamefont {Matysiak}, \citenamefont {Delle~Site},
  \citenamefont {Kremer},\ and\ \citenamefont
  {Clementi}}]{Praprotnik:2007:ARS}%
  \BibitemOpen
  \bibfield  {author} {\bibinfo {author} {\bibfnamefont {M.}~\bibnamefont
  {Praprotnik}}, \bibinfo {author} {\bibfnamefont {S.}~\bibnamefont
  {Matysiak}}, \bibinfo {author} {\bibfnamefont {L.}~\bibnamefont
  {Delle~Site}}, \bibinfo {author} {\bibfnamefont {K.}~\bibnamefont {Kremer}},
  \ and\ \bibinfo {author} {\bibfnamefont {C.}~\bibnamefont {Clementi}},\
  }\href@noop {} {\bibfield  {journal} {\bibinfo  {journal} {Journal of
  Physics: Condensed Matter}\ }\textbf {\bibinfo {volume} {19}},\ \bibinfo
  {pages} {292201} (\bibinfo {year} {2007})}\BibitemShut {NoStop}%
\bibitem [{\citenamefont {Ensing}\ \emph {et~al.}(2007)\citenamefont {Ensing},
  \citenamefont {Nielsen}, \citenamefont {Moore}, \citenamefont {Klein},\ and\
  \citenamefont {Parrinello}}]{Ensing:2007:ECA}%
  \BibitemOpen
  \bibfield  {author} {\bibinfo {author} {\bibfnamefont {B.}~\bibnamefont
  {Ensing}}, \bibinfo {author} {\bibfnamefont {S.}~\bibnamefont {Nielsen}},
  \bibinfo {author} {\bibfnamefont {P.}~\bibnamefont {Moore}}, \bibinfo
  {author} {\bibfnamefont {M.}~\bibnamefont {Klein}}, \ and\ \bibinfo {author}
  {\bibfnamefont {M.}~\bibnamefont {Parrinello}},\ }\href@noop {} {\bibfield
  {journal} {\bibinfo  {journal} {Journal of Chemical Theory and Computation}\
  }\textbf {\bibinfo {volume} {3}},\ \bibinfo {pages} {1100} (\bibinfo {year}
  {2007})}\BibitemShut {NoStop}%
\bibitem [{\citenamefont {Praprotnik}, \citenamefont {Delle~Site},\ and\
  \citenamefont {Kremer}(2005)}]{Praprotnik:2005:ARM}%
  \BibitemOpen
  \bibfield  {author} {\bibinfo {author} {\bibfnamefont {M.}~\bibnamefont
  {Praprotnik}}, \bibinfo {author} {\bibfnamefont {L.}~\bibnamefont
  {Delle~Site}}, \ and\ \bibinfo {author} {\bibfnamefont {K.}~\bibnamefont
  {Kremer}},\ }\href@noop {} {\bibfield  {journal} {\bibinfo  {journal}
  {Journal of Chemical Physics}\ }\textbf {\bibinfo {volume} {123}},\ \bibinfo
  {pages} {224106} (\bibinfo {year} {2005})}\BibitemShut {NoStop}%
\bibitem [{\citenamefont {Potestio}, \citenamefont {Peter},\ and\ \citenamefont
  {Kremer}(2014)}]{Potestio:2014:CSS}%
  \BibitemOpen
  \bibfield  {author} {\bibinfo {author} {\bibfnamefont {R.}~\bibnamefont
  {Potestio}}, \bibinfo {author} {\bibfnamefont {C.}~\bibnamefont {Peter}}, \
  and\ \bibinfo {author} {\bibfnamefont {K.}~\bibnamefont {Kremer}},\
  }\href@noop {} {\bibfield  {journal} {\bibinfo  {journal} {Entropy}\ }\textbf
  {\bibinfo {volume} {16}},\ \bibinfo {pages} {4199} (\bibinfo {year}
  {2014})}\BibitemShut {NoStop}%
\bibitem [{\citenamefont {Zavadlav}\ \emph {et~al.}(2014)\citenamefont
  {Zavadlav}, \citenamefont {Melo}, \citenamefont {Marrink},\ and\
  \citenamefont {Praprotnik}}]{Zavadlav:2014:ARS}%
  \BibitemOpen
  \bibfield  {author} {\bibinfo {author} {\bibfnamefont {J.}~\bibnamefont
  {Zavadlav}}, \bibinfo {author} {\bibfnamefont {M.}~\bibnamefont {Melo}},
  \bibinfo {author} {\bibfnamefont {S.}~\bibnamefont {Marrink}}, \ and\
  \bibinfo {author} {\bibfnamefont {M.}~\bibnamefont {Praprotnik}},\
  }\href@noop {} {\bibfield  {journal} {\bibinfo  {journal} {Journal of
  Chemical Physics}\ }\textbf {\bibinfo {volume} {140}},\ \bibinfo {pages}
  {054114} (\bibinfo {year} {2014})}\BibitemShut {NoStop}%
\bibitem [{\citenamefont {Zavadlav}, \citenamefont {Podgornik},\ and\
  \citenamefont {Praprotnik}(2015)}]{Zavadlav:2015:ARS}%
  \BibitemOpen
  \bibfield  {author} {\bibinfo {author} {\bibfnamefont {J.}~\bibnamefont
  {Zavadlav}}, \bibinfo {author} {\bibfnamefont {R.}~\bibnamefont {Podgornik}},
  \ and\ \bibinfo {author} {\bibfnamefont {M.}~\bibnamefont {Praprotnik}},\
  }\href@noop {} {\bibfield  {journal} {\bibinfo  {journal} {Journal of
  Chemical Theory and Computation}\ }\textbf {\bibinfo {volume} {11}},\
  \bibinfo {pages} {5035} (\bibinfo {year} {2015})}\BibitemShut {NoStop}%
\bibitem [{\citenamefont {Zavadlav}, \citenamefont {Bevc},\ and\ \citenamefont
  {Praprotnik}(2017)}]{Zavadlav:2017:ARS}%
  \BibitemOpen
  \bibfield  {author} {\bibinfo {author} {\bibfnamefont {J.}~\bibnamefont
  {Zavadlav}}, \bibinfo {author} {\bibfnamefont {S.}~\bibnamefont {Bevc}}, \
  and\ \bibinfo {author} {\bibfnamefont {M.}~\bibnamefont {Praprotnik}},\
  }\href@noop {} {\bibfield  {journal} {\bibinfo  {journal} {European
  Biophysics Journal}\ }\textbf {\bibinfo {volume} {46}},\ \bibinfo {pages}
  {821} (\bibinfo {year} {2017})}\BibitemShut {NoStop}%
\bibitem [{\citenamefont {Flegg}\ \emph {et~al.}(2014)\citenamefont {Flegg},
  \citenamefont {Chapman}, \citenamefont {Zheng},\ and\ \citenamefont
  {Erban}}]{Flegg:2014:ATM}%
  \BibitemOpen
  \bibfield  {author} {\bibinfo {author} {\bibfnamefont {M.}~\bibnamefont
  {Flegg}}, \bibinfo {author} {\bibfnamefont {J.}~\bibnamefont {Chapman}},
  \bibinfo {author} {\bibfnamefont {L.}~\bibnamefont {Zheng}}, \ and\ \bibinfo
  {author} {\bibfnamefont {R.}~\bibnamefont {Erban}},\ }\href@noop {}
  {\bibfield  {journal} {\bibinfo  {journal} {SIAM Journal on Scientific
  Computing}\ }\textbf {\bibinfo {volume} {36}},\ \bibinfo {pages} {B561}
  (\bibinfo {year} {2014})}\BibitemShut {NoStop}%
\bibitem [{\citenamefont {Flegg}, \citenamefont {Hellander},\ and\
  \citenamefont {Erban}(2015)}]{Flegg:2015:CMC}%
  \BibitemOpen
  \bibfield  {author} {\bibinfo {author} {\bibfnamefont {M.}~\bibnamefont
  {Flegg}}, \bibinfo {author} {\bibfnamefont {S.}~\bibnamefont {Hellander}}, \
  and\ \bibinfo {author} {\bibfnamefont {R.}~\bibnamefont {Erban}},\
  }\href@noop {} {\bibfield  {journal} {\bibinfo  {journal} {Journal of
  Computational Physics}\ }\textbf {\bibinfo {volume} {289}},\ \bibinfo {pages}
  {1} (\bibinfo {year} {2015})}\BibitemShut {NoStop}%
\bibitem [{\citenamefont {Robinson}, \citenamefont {Flegg},\ and\ \citenamefont
  {Erban}(2014)}]{Robinson:2014:ATM}%
  \BibitemOpen
  \bibfield  {author} {\bibinfo {author} {\bibfnamefont {M.}~\bibnamefont
  {Robinson}}, \bibinfo {author} {\bibfnamefont {M.}~\bibnamefont {Flegg}}, \
  and\ \bibinfo {author} {\bibfnamefont {R.}~\bibnamefont {Erban}},\
  }\href@noop {} {\bibfield  {journal} {\bibinfo  {journal} {Journal of
  chemical physics}\ }\textbf {\bibinfo {volume} {140}},\ \bibinfo {pages}
  {124109} (\bibinfo {year} {2014})}\BibitemShut {NoStop}%
\bibitem [{\citenamefont {Smith}\ and\ \citenamefont
  {Yates}(2018)}]{Smith:2018:SEH}%
  \BibitemOpen
  \bibfield  {author} {\bibinfo {author} {\bibfnamefont {C.}~\bibnamefont
  {Smith}}\ and\ \bibinfo {author} {\bibfnamefont {C.}~\bibnamefont {Yates}},\
  }\href@noop {} {\bibfield  {journal} {\bibinfo  {journal} {Journal of The
  Royal Society Interface}\ }\textbf {\bibinfo {volume} {15}},\ \bibinfo
  {pages} {20170931} (\bibinfo {year} {2018})}\BibitemShut {NoStop}%
\bibitem [{\citenamefont {Franz}\ \emph {et~al.}(2013)\citenamefont {Franz},
  \citenamefont {Flegg}, \citenamefont {Chapman},\ and\ \citenamefont
  {Erban}}]{Franz:2013:MRA}%
  \BibitemOpen
  \bibfield  {author} {\bibinfo {author} {\bibfnamefont {B.}~\bibnamefont
  {Franz}}, \bibinfo {author} {\bibfnamefont {M.}~\bibnamefont {Flegg}},
  \bibinfo {author} {\bibfnamefont {J.}~\bibnamefont {Chapman}}, \ and\
  \bibinfo {author} {\bibfnamefont {R.}~\bibnamefont {Erban}},\ }\href@noop {}
  {\bibfield  {journal} {\bibinfo  {journal} {SIAM Journal on Applied
  Mathematics}\ }\textbf {\bibinfo {volume} {73}},\ \bibinfo {pages} {1224}
  (\bibinfo {year} {2013})}\BibitemShut {NoStop}%
\bibitem [{\citenamefont {Delgado-Buscalioni}, \citenamefont {Kremer},\ and\
  \citenamefont {Praprotnik}(2009)}]{Delgado:2009:CAC}%
  \BibitemOpen
  \bibfield  {author} {\bibinfo {author} {\bibfnamefont {R.}~\bibnamefont
  {Delgado-Buscalioni}}, \bibinfo {author} {\bibfnamefont {K.}~\bibnamefont
  {Kremer}}, \ and\ \bibinfo {author} {\bibfnamefont {M.}~\bibnamefont
  {Praprotnik}},\ }\href@noop {} {\bibfield  {journal} {\bibinfo  {journal}
  {Journal of Chemical Physics}\ }\textbf {\bibinfo {volume} {131}},\ \bibinfo
  {pages} {244107} (\bibinfo {year} {2009})}\BibitemShut {NoStop}%
\bibitem [{\citenamefont {Machado}, \citenamefont {Gonz\'ales},\ and\
  \citenamefont {Pantano}(2017)}]{Machado:2017:MSV}%
  \BibitemOpen
  \bibfield  {author} {\bibinfo {author} {\bibfnamefont {M.}~\bibnamefont
  {Machado}}, \bibinfo {author} {\bibfnamefont {H.}~\bibnamefont {Gonz\'ales}},
  \ and\ \bibinfo {author} {\bibfnamefont {S.}~\bibnamefont {Pantano}},\
  }\href@noop {} {\bibfield  {journal} {\bibinfo  {journal} {Journal of
  Chemical Theory and Computation}\ }\textbf {\bibinfo {volume} {13}},\
  \bibinfo {pages} {5106–5116} (\bibinfo {year} {2017})}\BibitemShut
  {NoStop}%
\bibitem [{\citenamefont {Tarasova}\ \emph {et~al.}(2017)\citenamefont
  {Tarasova}, \citenamefont {Farafonov}, \citenamefont {Khayat}, \citenamefont
  {Okimoto}, \citenamefont {Komatsu}, \citenamefont {Taiji},\ and\
  \citenamefont {Nerukh}}]{Tarasova:2017:AMD}%
  \BibitemOpen
  \bibfield  {author} {\bibinfo {author} {\bibfnamefont {E.}~\bibnamefont
  {Tarasova}}, \bibinfo {author} {\bibfnamefont {V.}~\bibnamefont {Farafonov}},
  \bibinfo {author} {\bibfnamefont {R.}~\bibnamefont {Khayat}}, \bibinfo
  {author} {\bibfnamefont {N.}~\bibnamefont {Okimoto}}, \bibinfo {author}
  {\bibfnamefont {T.}~\bibnamefont {Komatsu}}, \bibinfo {author} {\bibfnamefont
  {M.}~\bibnamefont {Taiji}}, \ and\ \bibinfo {author} {\bibfnamefont
  {D.}~\bibnamefont {Nerukh}},\ }\href@noop {} {\bibfield  {journal} {\bibinfo
  {journal} {Journal of Physical Chemistry Letters}\ }\textbf {\bibinfo
  {volume} {8}},\ \bibinfo {pages} {779} (\bibinfo {year} {2017})}\BibitemShut
  {NoStop}%
\bibitem [{\citenamefont {Fogarty}, \citenamefont {Potestio},\ and\
  \citenamefont {Kremer}(2016)}]{Fogarty:2016:MMC}%
  \BibitemOpen
  \bibfield  {author} {\bibinfo {author} {\bibfnamefont {A.}~\bibnamefont
  {Fogarty}}, \bibinfo {author} {\bibfnamefont {R.}~\bibnamefont {Potestio}}, \
  and\ \bibinfo {author} {\bibfnamefont {K.}~\bibnamefont {Kremer}},\
  }\href@noop {} {\bibfield  {journal} {\bibinfo  {journal} {Proteins}\
  }\textbf {\bibinfo {volume} {84}},\ \bibinfo {pages} {1902} (\bibinfo {year}
  {2016})}\BibitemShut {NoStop}%
\bibitem [{\citenamefont {Rolls}, \citenamefont {Togashi},\ and\ \citenamefont
  {Erban}(2017)}]{Rolls:2017:VRR}%
  \BibitemOpen
  \bibfield  {author} {\bibinfo {author} {\bibfnamefont {E.}~\bibnamefont
  {Rolls}}, \bibinfo {author} {\bibfnamefont {Y.}~\bibnamefont {Togashi}}, \
  and\ \bibinfo {author} {\bibfnamefont {R.}~\bibnamefont {Erban}},\
  }\href@noop {} {\bibfield  {journal} {\bibinfo  {journal} {Multiscale
  Modeling and Simulation}\ }\textbf {\bibinfo {volume} {15}},\ \bibinfo
  {pages} {1672} (\bibinfo {year} {2017})}\BibitemShut {NoStop}%
\bibitem [{\citenamefont {Rolls}\ and\ \citenamefont
  {Erban}(2018)}]{Rolls:2018:MPB}%
  \BibitemOpen
  \bibfield  {author} {\bibinfo {author} {\bibfnamefont {E.}~\bibnamefont
  {Rolls}}\ and\ \bibinfo {author} {\bibfnamefont {R.}~\bibnamefont {Erban}},\
  }\href@noop {} {\bibfield  {journal} {\bibinfo  {journal} {Journal of
  Chemical Physics}\ }\textbf {\bibinfo {volume} {148}},\ \bibinfo {pages}
  {194111} (\bibinfo {year} {2018})}\BibitemShut {NoStop}%
\bibitem [{\citenamefont {Di~Pasquale}, \citenamefont {Gowers},\ and\
  \citenamefont {Carbone}(2014)}]{DiPasquale:2014:MTS}%
  \BibitemOpen
  \bibfield  {author} {\bibinfo {author} {\bibfnamefont {N.}~\bibnamefont
  {Di~Pasquale}}, \bibinfo {author} {\bibfnamefont {R.}~\bibnamefont {Gowers}},
  \ and\ \bibinfo {author} {\bibfnamefont {P.}~\bibnamefont {Carbone}},\
  }\href@noop {} {\bibfield  {journal} {\bibinfo  {journal} {Journal of
  Computational Chemistry}\ }\textbf {\bibinfo {volume} {35}},\ \bibinfo
  {pages} {1199} (\bibinfo {year} {2014})}\BibitemShut {NoStop}%
\bibitem [{\citenamefont {Di~Pasquale}\ and\ \citenamefont
  {Carbone}(2017)}]{DiPasquale:2017:LGD}%
  \BibitemOpen
  \bibfield  {author} {\bibinfo {author} {\bibfnamefont {N.}~\bibnamefont
  {Di~Pasquale}}\ and\ \bibinfo {author} {\bibfnamefont {P.}~\bibnamefont
  {Carbone}},\ }\href@noop {} {\bibfield  {journal} {\bibinfo  {journal}
  {Journal of Chemical Physics}\ }\textbf {\bibinfo {volume} {146}},\ \bibinfo
  {pages} {084905} (\bibinfo {year} {2017})}\BibitemShut {NoStop}%
\bibitem [{\citenamefont {Dama}\ \emph {et~al.}(2013)\citenamefont {Dama},
  \citenamefont {Sinitskiy}, \citenamefont {McCullagh}, \citenamefont {Weare},
  \citenamefont {Roux}, \citenamefont {Dinner},\ and\ \citenamefont
  {Voth}}]{Dama:2013:DUC}%
  \BibitemOpen
  \bibfield  {author} {\bibinfo {author} {\bibfnamefont {J.}~\bibnamefont
  {Dama}}, \bibinfo {author} {\bibfnamefont {A.}~\bibnamefont {Sinitskiy}},
  \bibinfo {author} {\bibfnamefont {M.}~\bibnamefont {McCullagh}}, \bibinfo
  {author} {\bibfnamefont {J.}~\bibnamefont {Weare}}, \bibinfo {author}
  {\bibfnamefont {B.}~\bibnamefont {Roux}}, \bibinfo {author} {\bibfnamefont
  {A.}~\bibnamefont {Dinner}}, \ and\ \bibinfo {author} {\bibfnamefont
  {G.}~\bibnamefont {Voth}},\ }\href@noop {} {\bibfield  {journal} {\bibinfo
  {journal} {Journal of Chemical Theory and Computation}\ }\textbf {\bibinfo
  {volume} {9}},\ \bibinfo {pages} {2466} (\bibinfo {year} {2013})}\BibitemShut
  {NoStop}%
\bibitem [{\citenamefont {Robert}(1995)}]{Robert:1995:STN}%
  \BibitemOpen
  \bibfield  {author} {\bibinfo {author} {\bibfnamefont {C.}~\bibnamefont
  {Robert}},\ }\href@noop {} {\bibfield  {journal} {\bibinfo  {journal}
  {Statistics and Computing}\ }\textbf {\bibinfo {volume} {5}},\ \bibinfo
  {pages} {121} (\bibinfo {year} {1995})}\BibitemShut {NoStop}%
\bibitem [{\citenamefont {van Gunsteren}\ and\ \citenamefont
  {Berendsen}(1982)}]{vanGunsteren:1982:ABD}%
  \BibitemOpen
  \bibfield  {author} {\bibinfo {author} {\bibfnamefont {W.}~\bibnamefont {van
  Gunsteren}}\ and\ \bibinfo {author} {\bibfnamefont {H.}~\bibnamefont
  {Berendsen}},\ }\href@noop {} {\bibfield  {journal} {\bibinfo  {journal}
  {Molecular Physics}\ }\textbf {\bibinfo {volume} {45}},\ \bibinfo {pages}
  {637} (\bibinfo {year} {1982})}\BibitemShut {NoStop}%
\bibitem [{\citenamefont {Wang}\ and\ \citenamefont
  {Skeel}(2003)}]{Wang:2003:AFN}%
  \BibitemOpen
  \bibfield  {author} {\bibinfo {author} {\bibfnamefont {W.}~\bibnamefont
  {Wang}}\ and\ \bibinfo {author} {\bibfnamefont {R.}~\bibnamefont {Skeel}},\
  }\href@noop {} {\bibfield  {journal} {\bibinfo  {journal} {Molecular
  Physics}\ }\textbf {\bibinfo {volume} {101}},\ \bibinfo {pages} {2149}
  (\bibinfo {year} {2003})}\BibitemShut {NoStop}%
\bibitem [{\citenamefont {Brunger}, \citenamefont {Brooks},\ and\ \citenamefont
  {Karplus}(1984)}]{Brunger:1984:SBC}%
  \BibitemOpen
  \bibfield  {author} {\bibinfo {author} {\bibfnamefont {A.}~\bibnamefont
  {Brunger}}, \bibinfo {author} {\bibfnamefont {C.}~\bibnamefont {Brooks}}, \
  and\ \bibinfo {author} {\bibfnamefont {M.}~\bibnamefont {Karplus}},\
  }\href@noop {} {\bibfield  {journal} {\bibinfo  {journal} {Chemical Physics
  Letters}\ }\textbf {\bibinfo {volume} {105}},\ \bibinfo {pages} {495 }
  (\bibinfo {year} {1984})}\BibitemShut {NoStop}%
\bibitem [{\citenamefont {Gronbech-Jensen}\ and\ \citenamefont
  {Farago}(2013)}]{Gronbech:SEV:2013}%
  \BibitemOpen
  \bibfield  {author} {\bibinfo {author} {\bibfnamefont {N.}~\bibnamefont
  {Gronbech-Jensen}}\ and\ \bibinfo {author} {\bibfnamefont {O.}~\bibnamefont
  {Farago}},\ }\href@noop {} {\bibfield  {journal} {\bibinfo  {journal}
  {Molecular Physics}\ }\textbf {\bibinfo {volume} {111}},\ \bibinfo {pages}
  {983} (\bibinfo {year} {2013})}\BibitemShut {NoStop}%
\bibitem [{\citenamefont {Burrage}, \citenamefont {Lenane},\ and\ \citenamefont
  {Lythe}(2007)}]{Burrage:2007:NMS}%
  \BibitemOpen
  \bibfield  {author} {\bibinfo {author} {\bibfnamefont {K.}~\bibnamefont
  {Burrage}}, \bibinfo {author} {\bibfnamefont {I.}~\bibnamefont {Lenane}}, \
  and\ \bibinfo {author} {\bibfnamefont {G.}~\bibnamefont {Lythe}},\
  }\href@noop {} {\bibfield  {journal} {\bibinfo  {journal} {SIAM Journal on
  Scientific Computing}\ }\textbf {\bibinfo {volume} {29}},\ \bibinfo {pages}
  {245} (\bibinfo {year} {2007})}\BibitemShut {NoStop}%
\bibitem [{\citenamefont {Davtyan}\ \emph {et~al.}(2015)\citenamefont
  {Davtyan}, \citenamefont {Dama}, \citenamefont {Voth},\ and\ \citenamefont
  {Andersen}}]{Davtyan:2015:DFM}%
  \BibitemOpen
  \bibfield  {author} {\bibinfo {author} {\bibfnamefont {A.}~\bibnamefont
  {Davtyan}}, \bibinfo {author} {\bibfnamefont {J.}~\bibnamefont {Dama}},
  \bibinfo {author} {\bibfnamefont {G.}~\bibnamefont {Voth}}, \ and\ \bibinfo
  {author} {\bibfnamefont {H.}~\bibnamefont {Andersen}},\ }\href@noop {}
  {\bibfield  {journal} {\bibinfo  {journal} {Journal of Chemical Physics}\
  }\textbf {\bibinfo {volume} {142}},\ \bibinfo {pages} {154104} (\bibinfo
  {year} {2015})}\BibitemShut {NoStop}%
\bibitem [{\citenamefont {Erban}(2016)}]{Erban:2016:CAM}%
  \BibitemOpen
  \bibfield  {author} {\bibinfo {author} {\bibfnamefont {R.}~\bibnamefont
  {Erban}},\ }\href@noop {} {\bibfield  {journal} {\bibinfo  {journal}
  {Proceedings of the Royal Society A}\ }\textbf {\bibinfo {volume} {472}},\
  \bibinfo {pages} {20150556} (\bibinfo {year} {2016})}\BibitemShut {NoStop}%
\bibitem [{\citenamefont {Davtyan}, \citenamefont {Voth},\ and\ \citenamefont
  {Andersen}(2016)}]{Davtyan:2016:DFM}%
  \BibitemOpen
  \bibfield  {author} {\bibinfo {author} {\bibfnamefont {A.}~\bibnamefont
  {Davtyan}}, \bibinfo {author} {\bibfnamefont {G.}~\bibnamefont {Voth}}, \
  and\ \bibinfo {author} {\bibfnamefont {H.}~\bibnamefont {Andersen}},\
  }\href@noop {} {\bibfield  {journal} {\bibinfo  {journal} {Journal of
  Chemical Physics}\ }\textbf {\bibinfo {volume} {145}},\ \bibinfo {pages}
  {224107} (\bibinfo {year} {2016})}\BibitemShut {NoStop}%
\bibitem [{\citenamefont {Leimkuhler}, \citenamefont {Noorizadeh},\ and\
  \citenamefont {Theil}(2009)}]{Leimkuhler:2009:GST}%
  \BibitemOpen
  \bibfield  {author} {\bibinfo {author} {\bibfnamefont {B.}~\bibnamefont
  {Leimkuhler}}, \bibinfo {author} {\bibnamefont {Noorizadeh}}, \ and\ \bibinfo
  {author} {\bibnamefont {Theil}},\ }\href@noop {} {\bibfield  {journal}
  {\bibinfo  {journal} {Journal of Statistical Physics}\ }\textbf {\bibinfo
  {volume} {135}},\ \bibinfo {pages} {261} (\bibinfo {year}
  {2009})}\BibitemShut {NoStop}%
\bibitem [{\citenamefont {Erban}\ and\ \citenamefont
  {Chapman}(2019)}]{Erban:2019:SMR}%
  \BibitemOpen
  \bibfield  {author} {\bibinfo {author} {\bibfnamefont {R.}~\bibnamefont
  {Erban}}\ and\ \bibinfo {author} {\bibfnamefont {S.~J.}\ \bibnamefont
  {Chapman}},\ }\href@noop {} {\emph {\bibinfo {title} {\hfill\break Stochastic modelling of
  reaction-diffusion processes\hfill\break}}}\ (\bibinfo  {publisher} {Cambridge University
  Press, ISBN 9781108498128},\ \bibinfo {year} {2019})\BibitemShut {NoStop}%
\bibitem [{\citenamefont {Miller}\ and\ \citenamefont
  {Tadmor}(2009)}]{Miller:UFP:2009}%
  \BibitemOpen
  \bibfield  {author} {\bibinfo {author} {\bibfnamefont {R.}~\bibnamefont
  {Miller}}\ and\ \bibinfo {author} {\bibfnamefont {E.}~\bibnamefont
  {Tadmor}},\ }\href@noop {} {\bibfield  {journal} {\bibinfo  {journal}
  {Modelling and Simulation in Materials Science and Engineering}\ }\textbf
  {\bibinfo {volume} {17}},\ \bibinfo {pages} {053001} (\bibinfo {year}
  {2009})}\BibitemShut {NoStop}%
\bibitem [{\citenamefont {Kevrekidis}\ \emph {et~al.}(2003)\citenamefont
  {Kevrekidis}, \citenamefont {Gear}, \citenamefont {Hyman}, \citenamefont
  {Kevrekidis}, \citenamefont {Runborg},\ and\ \citenamefont
  {Theodoropoulos}}]{Kevrekidis:2003:EFM}%
  \BibitemOpen
  \bibfield  {author} {\bibinfo {author} {\bibfnamefont {I.}~\bibnamefont
  {Kevrekidis}}, \bibinfo {author} {\bibfnamefont {C.}~\bibnamefont {Gear}},
  \bibinfo {author} {\bibfnamefont {J.}~\bibnamefont {Hyman}}, \bibinfo
  {author} {\bibfnamefont {P.}~\bibnamefont {Kevrekidis}}, \bibinfo {author}
  {\bibfnamefont {O.}~\bibnamefont {Runborg}}, \ and\ \bibinfo {author}
  {\bibfnamefont {K.}~\bibnamefont {Theodoropoulos}},\ }\href@noop {}
  {\bibfield  {journal} {\bibinfo  {journal} {Communications in Mathematical
  Sciences}\ }\textbf {\bibinfo {volume} {1}},\ \bibinfo {pages} {715}
  (\bibinfo {year} {2003})}\BibitemShut {NoStop}%
\bibitem [{\citenamefont {Biyikli}, \citenamefont {Yang},\ and\ \citenamefont
  {To}(2014)}]{Biyikli:2014:MMI}%
  \BibitemOpen
  \bibfield  {author} {\bibinfo {author} {\bibfnamefont {E.}~\bibnamefont
  {Biyikli}}, \bibinfo {author} {\bibfnamefont {Q.}~\bibnamefont {Yang}}, \
  and\ \bibinfo {author} {\bibfnamefont {A.}~\bibnamefont {To}},\ }\href@noop
  {} {\bibfield  {journal} {\bibinfo  {journal} {Computer Methods in Applied
  Mechanics and Engineering}\ }\textbf {\bibinfo {volume} {274}},\ \bibinfo
  {pages} {42 } (\bibinfo {year} {2014})}\BibitemShut {NoStop}%
\bibitem [{\citenamefont {Jung}, \citenamefont {Hanke},\ and\ \citenamefont
  {Schmid}(2017)}]{Jung:2017:IRM}%
  \BibitemOpen
  \bibfield  {author} {\bibinfo {author} {\bibfnamefont {G.}~\bibnamefont
  {Jung}}, \bibinfo {author} {\bibfnamefont {M.}~\bibnamefont {Hanke}}, \ and\
  \bibinfo {author} {\bibfnamefont {F.}~\bibnamefont {Schmid}},\ }\href@noop {}
  {\bibfield  {journal} {\bibinfo  {journal} {Journal of Chemical Theory and
  Computation}\ }\textbf {\bibinfo {volume} {13}},\ \bibinfo {pages} {2481}
  (\bibinfo {year} {2017})}\BibitemShut {NoStop}%
\bibitem [{\citenamefont {Shin}\ \emph {et~al.}(2010)\citenamefont {Shin},
  \citenamefont {Kim}, \citenamefont {Talkner},\ and\ \citenamefont
  {Lee}}]{Shin:2010:BMM}%
  \BibitemOpen
  \bibfield  {author} {\bibinfo {author} {\bibfnamefont {H.}~\bibnamefont
  {Shin}}, \bibinfo {author} {\bibfnamefont {C.}~\bibnamefont {Kim}}, \bibinfo
  {author} {\bibfnamefont {P.}~\bibnamefont {Talkner}}, \ and\ \bibinfo
  {author} {\bibfnamefont {E.}~\bibnamefont {Lee}},\ }\href@noop {} {\bibfield
  {journal} {\bibinfo  {journal} {Chemical Physics}\ }\textbf {\bibinfo
  {volume} {375}},\ \bibinfo {pages} {316} (\bibinfo {year}
  {2010})}\BibitemShut {NoStop}%
\bibitem [{\citenamefont {Erban}\ and\ \citenamefont
  {Chapman}(2009)}]{Erban:2009:SMR}%
  \BibitemOpen
  \bibfield  {author} {\bibinfo {author} {\bibfnamefont {R.}~\bibnamefont
  {Erban}}\ and\ \bibinfo {author} {\bibfnamefont {S.~J.}\ \bibnamefont
  {Chapman}},\ }\href@noop {} {\bibfield  {journal} {\bibinfo  {journal}
  {Physical Biology}\ }\textbf {\bibinfo {volume} {6}},\ \bibinfo {pages}
  {046001} (\bibinfo {year} {2009})}\BibitemShut {NoStop}%
\bibitem [{\citenamefont {Roberts}, \citenamefont {Stone},\ and\ \citenamefont
  {Luthey-Schulten}(2013)}]{Roberts:2013:LMH}%
  \BibitemOpen
  \bibfield  {author} {\bibinfo {author} {\bibfnamefont {E.}~\bibnamefont
  {Roberts}}, \bibinfo {author} {\bibfnamefont {J.}~\bibnamefont {Stone}}, \
  and\ \bibinfo {author} {\bibfnamefont {Z.}~\bibnamefont {Luthey-Schulten}},\
  }\href@noop {} {\bibfield  {journal} {\bibinfo  {journal} {Journal of
  Computational Chemistry}\ }\textbf {\bibinfo {volume} {34}},\ \bibinfo
  {pages} {245} (\bibinfo {year} {2013})}\BibitemShut {NoStop}%
\bibitem [{\citenamefont {Erban}, \citenamefont {Flegg},\ and\ \citenamefont
  {Papoian}(2014)}]{Erban:2014:MSR}%
  \BibitemOpen
  \bibfield  {author} {\bibinfo {author} {\bibfnamefont {R.}~\bibnamefont
  {Erban}}, \bibinfo {author} {\bibfnamefont {M.}~\bibnamefont {Flegg}}, \ and\
  \bibinfo {author} {\bibfnamefont {G.}~\bibnamefont {Papoian}},\ }\href@noop
  {} {\bibfield  {journal} {\bibinfo  {journal} {Bulletin of Mathematical
  Biology}\ }\textbf {\bibinfo {volume} {76}},\ \bibinfo {pages} {799}
  (\bibinfo {year} {2014})}\BibitemShut {NoStop}%
\bibitem [{\citenamefont {Dobramysl}, \citenamefont {Papoian},\ and\
  \citenamefont {Erban}(2016)}]{Dobramysl:2016:SEI}%
  \BibitemOpen
  \bibfield  {author} {\bibinfo {author} {\bibfnamefont {U.}~\bibnamefont
  {Dobramysl}}, \bibinfo {author} {\bibfnamefont {G.}~\bibnamefont {Papoian}},
  \ and\ \bibinfo {author} {\bibfnamefont {E.}~\bibnamefont {Erban}},\
  }\href@noop {} {\bibfield  {journal} {\bibinfo  {journal} {Biophysical
  Journal}\ }\textbf {\bibinfo {volume} {110}},\ \bibinfo {pages} {2066}
  (\bibinfo {year} {2016})}\BibitemShut {NoStop}%
\bibitem [{\citenamefont {Dobramysl}, \citenamefont {R\"udiger},\ and\
  \citenamefont {Erban}(2016)}]{Dobramysl:2016:PMM}%
  \BibitemOpen
  \bibfield  {author} {\bibinfo {author} {\bibfnamefont {U.}~\bibnamefont
  {Dobramysl}}, \bibinfo {author} {\bibfnamefont {S.}~\bibnamefont
  {R\"udiger}}, \ and\ \bibinfo {author} {\bibfnamefont {R.}~\bibnamefont
  {Erban}},\ }\href@noop {} {\bibfield  {journal} {\bibinfo  {journal}
  {Multiscale Modelling and Simulation}\ }\textbf {\bibinfo {volume} {14}},\
  \bibinfo {pages} {997} (\bibinfo {year} {2016})}\BibitemShut {NoStop}%
\bibitem [{\citenamefont {Allen}, \citenamefont {Kuyucak},\ and\ \citenamefont
  {Chung}(2000)}]{Allen:2000:MDE}%
  \BibitemOpen
  \bibfield  {author} {\bibinfo {author} {\bibfnamefont {T.}~\bibnamefont
  {Allen}}, \bibinfo {author} {\bibfnamefont {S.}~\bibnamefont {Kuyucak}}, \
  and\ \bibinfo {author} {\bibfnamefont {S.}~\bibnamefont {Chung}},\
  }\href@noop {} {\bibfield  {journal} {\bibinfo  {journal} {Biophysical
  Chemistry}\ }\textbf {\bibinfo {volume} {86}},\ \bibinfo {pages} {1}
  (\bibinfo {year} {2000})}\BibitemShut {NoStop}%
\bibitem [{\citenamefont {Jensen}\ \emph {et~al.}(2010)\citenamefont {Jensen},
  \citenamefont {Borhani}, \citenamefont {Lindorff-Larsen}, \citenamefont
  {Maragakis}, \citenamefont {Jogini}, \citenamefont {Eastwood}, \citenamefont
  {Dror},\ and\ \citenamefont {Shaw}}]{Jensen:2010:PCH}%
  \BibitemOpen
  \bibfield  {author} {\bibinfo {author} {\bibfnamefont {M.}~\bibnamefont
  {Jensen}}, \bibinfo {author} {\bibfnamefont {D.}~\bibnamefont {Borhani}},
  \bibinfo {author} {\bibfnamefont {K.}~\bibnamefont {Lindorff-Larsen}},
  \bibinfo {author} {\bibfnamefont {P.}~\bibnamefont {Maragakis}}, \bibinfo
  {author} {\bibfnamefont {V.}~\bibnamefont {Jogini}}, \bibinfo {author}
  {\bibfnamefont {M.}~\bibnamefont {Eastwood}}, \bibinfo {author}
  {\bibfnamefont {R.}~\bibnamefont {Dror}}, \ and\ \bibinfo {author}
  {\bibfnamefont {D.}~\bibnamefont {Shaw}},\ }\href@noop {} {\bibfield
  {journal} {\bibinfo  {journal} {Proceedings of the National Academy of
  Sciences USA}\ }\textbf {\bibinfo {volume} {107}},\ \bibinfo {pages} {5833}
  (\bibinfo {year} {2010})}\BibitemShut {NoStop}%
\bibitem [{\citenamefont {Erban}\ and\ \citenamefont
  {Chapman}(2007{\natexlab{a}})}]{Erban:2007:RBC}%
  \BibitemOpen
  \bibfield  {author} {\bibinfo {author} {\bibfnamefont {R.}~\bibnamefont
  {Erban}}\ and\ \bibinfo {author} {\bibfnamefont {S.~J.}\ \bibnamefont
  {Chapman}},\ }\href@noop {} {\bibfield  {journal} {\bibinfo  {journal}
  {Physical Biology}\ }\textbf {\bibinfo {volume} {4}},\ \bibinfo {pages} {16}
  (\bibinfo {year} {2007}{\natexlab{a}})}\BibitemShut {NoStop}%
\bibitem [{\citenamefont {Erban}\ and\ \citenamefont
  {Chapman}(2007{\natexlab{b}})}]{Erban:2007:TSR}%
  \BibitemOpen
  \bibfield  {author} {\bibinfo {author} {\bibfnamefont {R.}~\bibnamefont
  {Erban}}\ and\ \bibinfo {author} {\bibfnamefont {S.~J.}\ \bibnamefont
  {Chapman}},\ }\href@noop {} {\bibfield  {journal} {\bibinfo  {journal}
  {Physical Review E}\ }\textbf {\bibinfo {volume} {75}},\ \bibinfo {pages}
  {041116} (\bibinfo {year} {2007}{\natexlab{b}})}\BibitemShut {NoStop}%
\bibitem [{\citenamefont {Chapman}, \citenamefont {Erban},\ and\ \citenamefont
  {Isaacson}(2016)}]{Chapman:2016:RBC}%
  \BibitemOpen
  \bibfield  {author} {\bibinfo {author} {\bibfnamefont {J.}~\bibnamefont
  {Chapman}}, \bibinfo {author} {\bibfnamefont {R.}~\bibnamefont {Erban}}, \
  and\ \bibinfo {author} {\bibfnamefont {S.}~\bibnamefont {Isaacson}},\
  }\href@noop {} {\bibfield  {journal} {\bibinfo  {journal} {SIAM Journal on
  Applied Mathematics}\ }\textbf {\bibinfo {volume} {76}},\ \bibinfo {pages}
  {368} (\bibinfo {year} {2016})}\BibitemShut {NoStop}%
\bibitem [{\citenamefont {Delgado-Buscalioni}\ and\ \citenamefont
  {Praprotnik}(2015)}]{Delgado:2015:OBM}%
  \BibitemOpen
  \bibfield  {author} {\bibinfo {author} {\bibfnamefont {J.}~\bibnamefont
  {Delgado-Buscalioni}, \bibfnamefont {R.and~Sabli{\'{c}}}}\ and\ \bibinfo
  {author} {\bibfnamefont {M.}~\bibnamefont {Praprotnik}},\ }\href@noop {}
  {\bibfield  {journal} {\bibinfo  {journal} {European Physical Journal Special
  Topics}\ }\textbf {\bibinfo {volume} {224}},\ \bibinfo {pages} {2331}
  (\bibinfo {year} {2015})}\BibitemShut {NoStop}%
\bibitem [{\citenamefont {Zavadlav}\ \emph {et~al.}(2018)\citenamefont
  {Zavadlav}, \citenamefont {Sablic}, \citenamefont {Podgornik},\ and\
  \citenamefont {Praprotnik}}]{Zavadlav:2018:ARS}%
  \BibitemOpen
  \bibfield  {author} {\bibinfo {author} {\bibfnamefont {J.}~\bibnamefont
  {Zavadlav}}, \bibinfo {author} {\bibfnamefont {J.}~\bibnamefont {Sablic}},
  \bibinfo {author} {\bibfnamefont {R.}~\bibnamefont {Podgornik}}, \ and\
  \bibinfo {author} {\bibfnamefont {M.}~\bibnamefont {Praprotnik}},\
  }\href@noop {} {\bibfield  {journal} {\bibinfo  {journal} {Biophysical
  Journal}\ }\textbf {\bibinfo {volume} {114}},\ \bibinfo {pages} {2352}
  (\bibinfo {year} {2018})}\BibitemShut {NoStop}%
\bibitem [{\citenamefont {Espa\~nol}(2009)}]{Espanol:2009:HDC}%
  \BibitemOpen
  \bibfield  {author} {\bibinfo {author} {\bibfnamefont {P.}~\bibnamefont
  {Espa\~nol}},\ }\href@noop {} {\bibfield  {journal} {\bibinfo  {journal}
  {Europhysics Letters}\ }\textbf {\bibinfo {volume} {88}},\ \bibinfo {pages}
  {40008} (\bibinfo {year} {2009})}\BibitemShut {NoStop}%
\bibitem [{\citenamefont {Di~Pasquale}, \citenamefont {Hudson},\ and\
  \citenamefont {Icardi}(2018)}]{DiPasquale:2018:SDH}%
  \BibitemOpen
  \bibfield  {author} {\bibinfo {author} {\bibfnamefont {N.}~\bibnamefont
  {Di~Pasquale}}, \bibinfo {author} {\bibfnamefont {T.}~\bibnamefont {Hudson}},
  \ and\ \bibinfo {author} {\bibfnamefont {M.}~\bibnamefont {Icardi}},\
  }\href@noop {} {\enquote {\bibinfo {title} {Systematic derivation of hybrid
  coarse-grained models},}\ } (\bibinfo {year} {2018}),\ \bibinfo {note}
  {available as https://arxiv.org/abs/1804.08157}\BibitemShut {NoStop}%
\bibitem [{\citenamefont {Zavadlav}\ and\ \citenamefont
  {Praprotnik}(2017)}]{Zavadlavb:2017:ARS}%
  \BibitemOpen
  \bibfield  {author} {\bibinfo {author} {\bibfnamefont {J.}~\bibnamefont
  {Zavadlav}}\ and\ \bibinfo {author} {\bibfnamefont {M.}~\bibnamefont
  {Praprotnik}},\ }\href@noop {} {\bibfield  {journal} {\bibinfo  {journal}
  {Journal of Chemical Physics}\ }\textbf {\bibinfo {volume} {147}},\ \bibinfo
  {pages} {114110} (\bibinfo {year} {2017})}\BibitemShut {NoStop}%
\bibitem [{\citenamefont {Gonzalez}, \citenamefont {Darr\'e},\ and\
  \citenamefont {Pantano}(2013)}]{Gonzalez:2013:TMA}%
  \BibitemOpen
  \bibfield  {author} {\bibinfo {author} {\bibfnamefont {H.}~\bibnamefont
  {Gonzalez}}, \bibinfo {author} {\bibfnamefont {L.}~\bibnamefont {Darr\'e}}, \
  and\ \bibinfo {author} {\bibfnamefont {S.}~\bibnamefont {Pantano}},\
  }\href@noop {} {\bibfield  {journal} {\bibinfo  {journal} {Journal of
  Physical Chemistry B}\ }\textbf {\bibinfo {volume} {117}},\ \bibinfo {pages}
  {14438} (\bibinfo {year} {2013})}\BibitemShut {NoStop}%
\bibitem [{\citenamefont {Liu}\ \emph {et~al.}(2012)\citenamefont {Liu},
  \citenamefont {Pu}, \citenamefont {Li}, \citenamefont {Shaffer},
  \citenamefont {Hoops}, \citenamefont {Tyson},\ and\ \citenamefont
  {Cao}}]{Liu:2012:HMS}%
  \BibitemOpen
  \bibfield  {author} {\bibinfo {author} {\bibfnamefont {Z.}~\bibnamefont
  {Liu}}, \bibinfo {author} {\bibfnamefont {Y.}~\bibnamefont {Pu}}, \bibinfo
  {author} {\bibfnamefont {F.}~\bibnamefont {Li}}, \bibinfo {author}
  {\bibfnamefont {C.}~\bibnamefont {Shaffer}}, \bibinfo {author} {\bibfnamefont
  {S.}~\bibnamefont {Hoops}}, \bibinfo {author} {\bibfnamefont
  {J.}~\bibnamefont {Tyson}}, \ and\ \bibinfo {author} {\bibfnamefont
  {Y.}~\bibnamefont {Cao}},\ }\href@noop {} {\bibfield  {journal} {\bibinfo
  {journal} {Journal of Chemical Physics}\ }\textbf {\bibinfo {volume} {136}},\
  \bibinfo {pages} {034105} (\bibinfo {year} {2012})}\BibitemShut {NoStop}%
\bibitem [{\citenamefont {Duncan}, \citenamefont {Erban},\ and\ \citenamefont
  {Zygalakis}(2016)}]{Duncan:2016:HFS}%
  \BibitemOpen
  \bibfield  {author} {\bibinfo {author} {\bibfnamefont {A.}~\bibnamefont
  {Duncan}}, \bibinfo {author} {\bibfnamefont {R.}~\bibnamefont {Erban}}, \
  and\ \bibinfo {author} {\bibfnamefont {K.}~\bibnamefont {Zygalakis}},\
  }\href@noop {} {\bibfield  {journal} {\bibinfo  {journal} {Journal of
  Computational Physics}\ }\textbf {\bibinfo {volume} {326}},\ \bibinfo {pages}
  {398} (\bibinfo {year} {2016})}\BibitemShut {NoStop}%
\bibitem [{\citenamefont {Franz}\ and\ \citenamefont
  {Erban}(2012)}]{Franz:2012:HMI}%
  \BibitemOpen
  \bibfield  {author} {\bibinfo {author} {\bibfnamefont {B.}~\bibnamefont
  {Franz}}\ and\ \bibinfo {author} {\bibfnamefont {R.}~\bibnamefont {Erban}},\
  }in\ \href@noop {} {\emph {\bibinfo {booktitle} {Dispersal, individual
  movement and spatial ecology: A mathematical perspective}}},\ \bibinfo
  {editor} {edited by\ \bibinfo {editor} {\bibfnamefont {M.}~\bibnamefont
  {Lewis}}, \bibinfo {editor} {\bibfnamefont {P.}~\bibnamefont {Maini}}, \ and\
  \bibinfo {editor} {\bibfnamefont {S.}~\bibnamefont {Petrovskii}}}\ (\bibinfo
  {publisher} {Springer},\ \bibinfo {year} {2012})\BibitemShut {NoStop}%
\end{thebibliography}
\end{document}